\documentclass[prl,twocolumn,longbibliography,superscriptaddress]{revtex4-1}
\usepackage{graphicx}
\usepackage{color}
\usepackage{amssymb}
\usepackage{amsmath}
\usepackage[squaren]{SIunits}
\usepackage[caption=false]{subfig}
\include{NewCommands}

\usepackage{hyperref}

\definecolor{mygreen}{rgb}{0,0.5,0}
\definecolor{myred}{rgb}{0.75,0,0}
\definecolor{myblue}{rgb}{0,0,0.75}
\definecolor{mymagenta}{cmyk}{0,1,0,0.12}
\definecolor{mycyan}{cmyk}{1,0,0,0.12}
\definecolor{myorange}{rgb}{1,0.5,0}

\newcommand{\nuT}{\nu_T}

\newcommand{\nubin}{\nu_{\rm bin}}
\newcommand{\Sbar}{{\bar{S}}}

\newcommand{\Nave}{N_{\rm ave}}
\newcommand{\cov}{{\rm cov}}
\newcommand{\var}{{\rm var}}

\newcommand{\Aeff}{A_{\rm eff}}

\newcommand{\Lcell}{L_{\rm cell}}

\newcommand{\Eph}{E_{\rm ph}}
\newcommand{\omegalight}{\omega_{\rm light}}
\renewcommand{\omegalight}{\omega}

\newcommand{\PRLsection}[1]{\noindent \textit{#1} ---}
\newcommand{\GamTh}{\Gamma^{\rm th}}
\newcommand{\GamExp}{\Gamma^{\rm exp}}
\newcommand{\GamMLE}{\Gamma^{\rm MLE}}

\begin{document}

\title{
%Sensitivity limits and quantum enhancement of noise spectroscopies\\
Sensitivity, quantum limits,  and quantum enhancement of noise spectroscopies\\
%Sensitivity of noise spectroscopy and quantum enhancement thereof
}

\newcommand{\ICFO}
{
\affiliation{ICFO-Institut de Ciencies Fotoniques, The Barcelona Institute of Science and Technology, 08860 Castelldefels (Barcelona), Spain}}
\newcommand{\ICREA}
{
\affiliation{ICREA -- Instituci\'o Catalana de Recerca i Estudis
Avan\c{c}ats, 08015 Barcelona, Spain}
}
\newcommand{\UBelgrade}
{
\affiliation{Faculty of Physics, University of Belgrade, Studentski Trg 12-16, 11000 Belgrade, Serbia}
}
\newcommand{\ECNU}
{
\affiliation{East China Normal University, Shanghai 200062,
China}
}

\author{Vito Giovanni Lucivero}
\ICFO
\email[Corresponding author: ]{lucivero@princeton.edu }
\altaffiliation{Current address: Department of Physics, Princeton University, Princeton, New Jersey 08544, USA}
\author{Aleksandra Dimic}
\ICFO
\UBelgrade
\author{Jia Kong}
\ICFO
\author{Ricardo Jim\'{e}nez-Mart\'{i}nez}
\ICFO
%\author{authors}
%\ICFO
\author{Morgan W. Mitchell}
\ICFO
\ICREA

\date{\today}

\begin{abstract}
We study the fundamental limits of noise spectroscopy using estimation theory, Faraday rotation probing of an atomic spin system, and squeezed light. We find a simple and general expression for the Fisher information, which quantifies the sensitivity to spectral parameters such as resonance frequency and linewidth. For optically-detected spin noise spectroscopy, we find that shot noise imposes ``local'' standard quantum limits for any given probe power and atom number, and also ``global'' standard quantum limits when probe power and atom number are taken as free parameters.  We confirm these estimation theory results using non-destructive Faraday rotation  probing of hot Rb vapor, observing the predicted optima and finding good quantitative agreement with a first-principles calculation of the spin noise spectra.  Finally, we show sensitivity beyond the atom- and photon-number-optimized global standard quantum limit using squeezed light.
\end{abstract}

\maketitle

\newcommand{\bg}{S_{\theta}}
\newcommand{\amp}{S_{\rm A}}
\newcommand{\area}{A}
\newcommand{\gam}{\gamma}
\newcommand{\nuL}{\nu_{\rm L}}
\newcommand{\nures}{\nu_{0}}
\newcommand{\conv}{\otimes}
\newcommand{\bigconv}{\bigotimes}
\newcommand{\supin}{^{(\rm in)}}
\newcommand{\supout}{^{(\rm out)}}
\newcommand{\xysig}{\sigma}
\newcommand{\Ssig}{\rho}

{Noise spectroscopies, in which naturally occurring fluctuations of a system of interest are recorded by a non-invasive probe, have applications in {a wide range of disciplines including} atomic \cite{Crooker2004} and solid state physics \cite{SikulaMIEL2002,VitusevichJSM2009}, surface science \cite{RomachPRL2015},  cell biology \cite{TolicNorrelykkePRL2004,TaylorNJP2013}, molecular biophysics \cite{KawakamiL2004,Berg-SorensenRSI2004}, geophysics \cite{DescherevskyNHESS2003},   space science \cite{MoncuquetGRL2005}, quantum opto-mecanics \cite{SafaviPRL2012}, and quantum information processing \cite{BiercukN2009,AlvarezPRL2011,YugePRL2011,BylanderNPhys2011,MedfordPRL2012}. % \btext{to name but a few}.
By the fluctuation-dissipation theorem, the noise spectrum under thermal equilibrium gives the same information as do driven spectroscopies, with the advantage of characterizing the system in its natural, undisturbed state \cite{KatsoprinakisPRA2007}.
%Because the signal is necessarily smaller than in driven spectroscopies, intrinsic statistical fluctuations \btext{ in the target system and probe} such as shot noise can play an important role in determining the sensitivity of such techniques \cite{Glasenapp2013,Lucivero2016}.
%\btext{Previous theoretical studies have estimated fundamental bounds to the precision of the technique in specific systems, such as optical tweezers \cite{TaylorNJP2013}, however their use in more general scenarios remains unclear. Of particular interest are spin systems.}
Understanding the statistical sensitivity of noise spectroscopy is essential for rigorous use of the technique in any of these fields. We study this problem from the perspective of parameter estimation theory, to derive the covariance matrix for spectral parameters obtained by fitting experimental spectra.

We illustrate and test the results using spin noise spectroscopy (SNS), a versatile technique that measures magnetic resonance features from thermal spin fluctations \cite{Zapasskii2013a}.  Non-optical SNS based on resonance force microscopy \cite{Rugar2004,Budakian2005} and NV-center magnetometry \cite{Balasubramanian2008,Maze2008,Mamin2012} have recently emerged, but still the most widely used technique is optical Faraday rotation (FR) to detect spin orientation \cite{Zap1981}. This FR-SNS is used to study spin physics in atomic gases \cite{Crooker2004,KatsoprinakisPRA2007} as well as conduction electrons \cite{Oestreich2005} and localized states in semiconductors \cite{Crooker2010,Glasenapp2013}. Extensions of SNS include measurements of cross-correlations of heterogeneous spin systems \cite{Dellis2014,Roy2015}, spin dynamics beyond thermal equilibrium \cite{Glasenapp2014} and multidimensional SNS \cite{Cundiff2013,KryvohuzJCP2015}, see \cite{Sinitsyn2016} for a review.

Quantum statistical fluctuations such as shot noise are often limiting in noise spectroscopies \cite{TaylorNJP2013,Glasenapp2013,Lucivero2016}, making it important to understand quantum limits and techniques to overcome them.  A general framework to estimate the spectra of noisy classical forces influencing quantum systems has been applied to spectroscopy by homodyne detection of an externally-imposed noisy phase \cite{NgPRA2016}. This framework provides fundamental limits but can be applied to noise spectroscopies only in the weak probing regime, due to the assumed ``classical,'' i.e., imperturbable, nature of the estimated force.

In contrast, our results make no classicality assumption. For FR-SNS we show two new results concerning the  quantum limits of the technique: first, availability of unlimited particle-number resources gives rise to an optimal sensitivity at finite number, in contrast to standard models from quantum metrology \cite{GiovannettiS2004,LangPRL2013}, which have sensitivity monotonic in particle number and thus must assume an externally-imposed constraint to give meaningful results, and second, the number-optimized standard-quantum-limit sensitivity can be surpassed using  squeezed-light probes. These theoretical predictions are tested by comparison against FR-SNS of hot rubidium vapor using a quantum-noise-limited probing system \cite{Lucivero2016} and atom-resonant optical squeezing \cite{PredojevicPRA2008, WolfgrammPRL2010}. The techniques demonstrated here enable \textit{a priori} optimization of experimental conditions, to maximize the information obtained from noise spectra.

\newcommand{\BinxAve}{{\cal N}}
\newcommand{\Nbin}{N_{\rm bin}}

\PRLsection{Sensitivity to spectral parameters} First we derive the covariance matrix for parameters obtained from noise spectra, by using their statistical properties and estimation theory. Noise spectroscopies record an observable $Y(t)$ %in FR a Stokes parameter of the probe beam,
that carries information about the physical system of interest \cite{MullerAPL2010}. Statistically, $Y(t)$ is a stationary random process with the long-time power spectral density $f(\nu,\mathbf{v})$, where $f$ describes a family of possible spectra, $\nu$ is the linear frequency and $\mathbf{v}$ is a vector of unknown parameters. %To estimate $\mathbf{v}$, w e sample
 $Y(t)$ is sampled at times $t_m = m \Delta$, $m \in \mathbb{N}$ for a time $T$, to obtain the discrete power spectrum    $S_i \equiv S(\nu_i) = |\tilde{Y}(\nu_i) |^2$, where $\tilde{Y}(\nu_i)$ is the discrete Fourier transform of $Y(t_m)$.
 % and is computed by discrete Fourier transform
%at times $t_i = i \Delta, i = 1\ldots T/\Delta$, from which
%the discrete spectrum
 %\equiv |\tilde{Y}(\nu_i)|^2$, where $\tilde{Y}(\nu_i)$ is the is computed
% of $Y(t_i)$,
Here $\nu_i = i \nuT$ are discrete frequencies with separation $\nuT \equiv 1/T$. When the acquisition is long relative to the coherence time of the noise process, so that the features of the spectrum are broad relative to $\nuT$,
%``stationary process long observation time'' (SPLOT)
we can take $\cov(S_i,S_j) = 0$
%for $\tilde{Y}(\nu_i), \tilde{Y}(\nu_j)$ are independent
for $i \ne j$ \cite{RootAMS1955,HamidiIEEETIT1975}, while ${\rm var}(S_i) = \langle S_i \rangle^2$.
%${\rm cov}(S_i,S_j) \equiv \langle S_i S_j \rangle - \langle S_i \rangle \langle S_j \rangle = \langle S_i \rangle^2 \delta_{ij}$,
%where $\delta$ is the Kronecker symbol. %Eq.~(\ref{Eq:covarS})
% .  For $i \ne 0$, stationarity implies $\langle \tilde{Y}(\nu_i) \rangle = 0$
%\btext{rjm: Why? do you mean $\langle \tilde{Y}(\nu_i) \rangle = 0? $ $ S_i$ is a positive-valued number, no? MWM: yes, thanks.  Do we need $\langle S_i \rangle$ for something ?  need to look at this. }
%\ctext{stationarity gives the Kronecker delta, however the squared mean is given by the fact that $S_i$ is described by an exponential distribution, should we mention this?}
% \begin{equation}
% \label{Eq:covarS}
%{\rm cov}(S_i,S_j) = \langle S_i \rangle^2 \delta_{ij}
%\end{equation}
As derived in the Supplementary Information (SI), this describes a unit signal-to-noise ratio $\langle S_i \rangle^2/\var(S_i) =1$ for $S_i$, independent of its true value, in marked contrast to most physical estimation problems.
%\btext{The signal-to-noise ratio ${\rm SNR} \equiv \langle S_i \rangle^2/\var(S_i)$ for $S_i$ is then unity, independent of the true value of $S_i$}, in marked contrast to most physical estimation problems.
%A detailed consideration shows that ... [distribution for $S_i$ ? would say it is exponential. ]

Two averaging procedures are often applied to reduce the uncertainty of {$S_i$, and thus increase SNR}:  a simple averaging of $\Nave$ independently acquired spectra $S^{(k)}$, and a ``coarse-grained'' averaging of $\Nbin \equiv \nubin/\nu_T$ adjacent values, giving the coarse-grained spectrum
$\Sbar_i \equiv (\Nbin\Nave)^{-1}\sum_{j=1}^{\Nbin} \sum_{k=1}^{\Nave} S_{i\Nbin+j}^{(k)}$.
%adjacent values to obtain %$\Sbar_j \equiv (S_{\Nbin j +1} + \ldots + S_{\Nbin j + \Nbin})/\Nbin$
%the coarse-grained spectrum
%$\Sbar_i \equiv \Nbin^{-1}\sum_{j=1}^{\Nbin} S_{i\Nbin+j}$ \ctext{rjm:this only includes the bin average, we need to include the average over $\Nave$ traces: $\Sbar_i \equiv \Nbin^{-1}\sum_{j=1}^{\Nbin} \Nave^{-1}\sum_{k=1}^{\Nave} S_{i\Nbin+j}^{(k)}$} \mtext{need to make this decimation agree with what was actually done. this is asymmetric about $\nu_i$ Was the real decimation asymmetric ? }. \ctext{I think the decimation was asymmetric, one option is: $\langle S_i \rangle _{\Nbin} \equiv \Nbin^{-1}\sum_{j=-\Nbin/2+1}^{\Nbin/2} S_{i+j}$ (this is the bin average only, extension to trace average is direct), where $\Nbin$ is an odd integer. Need to check what was actually done and how the decimated frequencies are defined.}
This averaging of $\BinxAve \equiv \Nbin\Nave$ uncorrelated contributions gives
%\begin{equation}
%\label{Eq:VarSBar}
%\var  \Sbar_i  = \frac{1}{\Nave  \Nbin } \langle \Sbar_i \rangle^2 \equiv \frac{1}{\BinxAve} \langle \Sbar_i \rangle^2.
%\end{equation}
$\var  (\Sbar_i)  =   \langle \Sbar_i \rangle^2/{\BinxAve}$.
By the central limit theorem, with increasing $\BinxAve$ the distribution of $\Sbar_i$ rapidly approaches a multivariate normal distribution with mean $\mu_i \equiv \langle \bar{S}_i \rangle =  f(\nu_i, {\bf v}) \equiv f_i$ and covariance matrix
\begin{equation}
\label{eq:Sigma}
\Sigma_{ij} \equiv \langle (\bar{S}_i-\mu_i)(\bar{S}_j-\mu_j) \rangle = \BinxAve^{-1}  f^2_i \delta_{ij}.
\end{equation}
%Te central limit theorem implies that
%the distribution of $\Sbar_i$ rapidly approaches normality with increasing $\BinxAve$. Asymptotically, the spectrum is drawn from a
%$\Sbar_i$ follows a multivariate normal distribution with mean $\mu_i \equiv \langle \bar{S}_i \rangle =  f(\nu_i, {\bf v}) \equiv f_i$ and covariance matrix \ctext{(rjm: here we use $\Sigma$ to represent the covariance matrix, in the rest of the paper we use $\Gamma$, why? to distinguish theory from experiment? I do not think so, can we stick with one symbol?) MWM: $\Sigma$ is for the covariance matrix of $S_i$, $\Gamma$ is the covariance matrix for ${v_i}$.  We need different symbols, but maybe they should be $\Gamma^{({v})}$ and $\Gamma^{({S})}$ ? That's not so convenient because we need at some point $\Sigma^{-1}$. }
%$\Sigma_{ij} \equiv \langle (\bar{S}_i-\mu_i)(\bar{S}_j-\mu_j) \rangle = \BinxAve^{-1}  f^2_i \delta_{ij}$.
%\begin{equation}
%\label{eq:Sigma}
%\Sigma_{ij} \equiv \langle (\bar{S}_i-\mu_i)(\bar{S}_j-\mu_j) \rangle = \BinxAve^{-1}  f^2_i \delta_{ij}.
%\end{equation}

Due to this normality, $\hat{\mathbf{v}}$ the maximum-likelihood estimator (MLE) of $\mathbf{v}$ is found by a fit that minimizes  $\chi^2 \equiv \sum_i (1-\bar{S}_i/f_i)^2$ by choice of $\mathbf{v}$.
%\btext{rjm: OK I got this.The information I was looking for is stated above.  Ignore the following comment:} \ctext{I need to go back to refresh my background on this. Some features of our problem are implicit in the exposition here, I feel it would be good for the reader if we explicitly mention them, I am not sure how to do it in an efficient manner.}
% is easily found by minimizing $\chi^2 \equiv \sum_i (1-\bar{S}_i/f_i)^2$.
%In the asymptotic regime, i.e., for large $\BinxAve$,  $\hat{\mathbf{v}}$
For large $\BinxAve$, this estimate saturates the Cramer-Rao bound \cite{LevinIEEETIT1965}, so that the error covariance matrix $\Gamma_{ab} \equiv \langle (\hat{\mathbf{v}}_a - \mathbf{v}_a)(\hat{\mathbf{v}}_b - \mathbf{v}_b) \rangle$ is simply $\Gamma = {\cal I}^{-1}$, where ${\cal I}$ is the Fisher information matrix for a vector-parametrized multivariate normal distribution \cite{KleinLAA2000}:
%Writing $\hat{\mathbf{v}}$ for the maximum-likelihood estimator (MLE) of $\mathbf{v}$, and $\Gamma_{ij} $ for the covariance matrix of $\hat{\mathbf{v}}$, the Cramer-Rao bound is $\Gamma_{} \ge {\cal I}^{-1}_{}$, where the Fisher information is \cite{KleinLAA2000}
\begin{eqnarray}
\label{Eq:CRBound}
{\cal I}_{jk} & = & \sum_i (\partial_j \mu_i) \Sigma^{-1} (\partial_k \mu_i)+ \frac{1}{2} {\rm Tr} \left[ \Sigma^{-1} (\partial_j \Sigma ) \Sigma^{-1} (\partial_k \Sigma)\right] \nonumber \\
& = & \BinxAve \sum_i (\partial_j f_i) f_i^{-2} (\partial_k f_i) + \frac{1}{2} \sum_i f_i^{-4} (\partial_j f_i^2)  (\partial_k f_i^2) \nonumber \\
& = & (\BinxAve+2) \sum_{i} f_i^{-2} (\partial_j f_i  )( \partial_k f_i),
%& = & (\BinxAve+2) \sum_{i}  (\partial_j \ln f_i  )( \partial_k \ln f_i),
\end{eqnarray}
%The last step is valid in the regime of many averages.
where $\partial_i$ indicates $\partial/\partial v_i$ and the second line follows from Eq.~(\ref{eq:Sigma}). An expansive derivation is given in the SI.

We note that the width of the bins, which affects both $\BinxAve$ and the number of terms in  $\sum_i$, does not alter ${\cal I}$, provided the  graining is not so coarse as to blur the spectral features. This justifies treating the sums as integrals.  We arrive to our first main result, the statistical sensitivity of spectral parameter estimation by noise spectroscopy:
%\begin{eqnarray}
%{\cal I}_{jk}
%& = & (\BinxAve+2) \sum_{i} f_i^{-2} (\partial_j f_i  )( \partial_k f_i) \nonumber \\
%& \approx &  {(\BinxAve+2)} \int_{\nu_1}^{\nu_2} \frac{d\nu}{\nu_T}  \frac{(\partial_j f_i  )( \partial_k f_i) }{f_i^{2}},
%\label{eq:Sensitivity}
%\end{eqnarray}
\begin{eqnarray}
%\Gamma & \equiv  & \langle (\hat{\mathbf{v}} - \mathbf{v})\wedge (\hat{\mathbf{v}} - \mathbf{v}) \rangle =  {\cal I}^{-1}
%\Gamma & =  &  {\cal I}^{-1} \nonumber \\
{\cal I}_{jk}
& = & (\BinxAve+2) \sum_{i}  (\partial_j \ln f_i  )( \partial_k \ln f_i) \nonumber \\
%& = &  (M+2) \sum_{i} f_i^{-2} (\partial_j f_i  )( \partial_l f_i) \nonumber \\
& \approx &  \frac{(\BinxAve+2)}{\nuT} \int_{\nu_1}^{\nu_2} {d\nu} \, {(\partial_j \ln f_i  )( \partial_k \ln f_i) }
%(\partial_j f_i  )( \partial_l f_i) f_i^{-2}
\label{eq:Sensitivity}
\end{eqnarray}
where ${\nu_1,\nu_2}$ delimit the frequency range over which the fit is performed \footnote{A similar result is given in Eq.~(2.32) of \cite{NgPRA2016}. Two important points of difference:  The expression in \cite{NgPRA2016} describes the \textit{quantum Fisher information}, i.e., the information available under the best possible measurement, while ours indicates the Fisher information \textit{per se}, the information of the measurement actually made. Also, our result involves the parameter dependence of $f$, including noise due to the probe, whereas that of \cite{NgPRA2016}  includes only $S_X$, the noise of the external force.}. Because noise spectra are used in many areas of physics, as well as biology and geosciences, Eq. (\ref{eq:Sensitivity}) can have wide usage.
%\ctext{We need to relate our work to that of Mankei Tsang \cite{Shilin2016}. I need to think about this.}  \mtext{MWM: how about the footnote I added ?}

%This is the statistical sensitivity of spectral parameter estimation by noise spectroscopy.
%Eq.~(\ref{Eq:CRBound}) and its approximation Eq.~(\ref{eq:IntegralApproximation}) describe the statistical limit of any noise spectroscopy.
%This is our first main result, the statistical sensitivity of noise spectroscopies for spectral parameter estimation.

\begin{figure}[t]
\centering
  \hspace{0.4cm}
\includegraphics[width=0.95 \columnwidth]{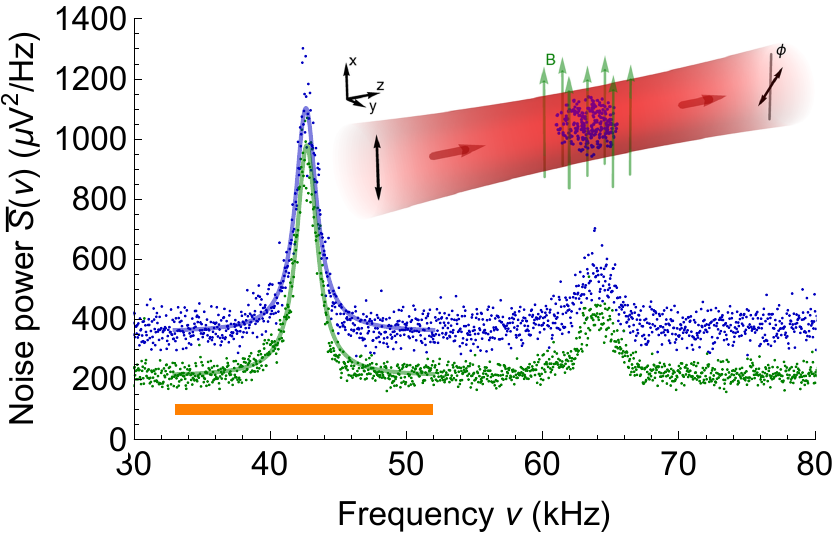}
  \caption{Representative noise spectra showing spin noise resonances of $^{85}$Rb at \unit{42.6}{\kilo\hertz} and $^{87}$Rb at \unit{64.0}{\kilo\hertz}.  Upper spectrum (blue) with coherent-state probing, lower spectrum (green) with polarization-squeezed probe light. Each spectrum is an average of $\Nave = 5$ acquisitions with coarse-graining of $\Nbin = 20$.  Data acquired with $n = \unit{4.23\times 10^{12}}{\centi\meter^{-3}}$ and $P = \unit{2.0}{\milli\watt}$. It is clearly seen that the noisiness $\var[ \Sbar(\nu) ]$ of the spectrum increases with increasing $\langle \Sbar(\nu) \rangle$. The squeezed probe  reduces the shot-noise background, with a beneficial effect on both the signal-to-noise ratio and the precision of spectral parameter estimation. Orange bar below spectra shows  fit region, blue and green curves show fits of Eq. (\ref{eq:fitfunction}) to the coherent and squeezed spectra, respectively. Inset: principle of spin noise measurement. Polarized light experiences Faraday rotation by an angle $\phi$ proportional to the on-axis magnetization of the atomic ensemble, and is detected with a polarimeter (not shown).
}
  \label{fig:Setup}
\end{figure}

\PRLsection{Validation}
%Application of the above results requires that $f$, the family of possible spectra, include the spectrum of the system under study.
To test the applicability of this result, we analyze {atomic vapor spin noise detected by} FR-SNS {in hot atomic vapor} (see  Fig. \ref{fig:Setup}), using the experimental system described in \cite{Lucivero2016}. In this case, the spectrum $f(\nu, {\bf v}) $ is
\begin{equation}
f(\nu, {\bf v}) = S_{\rm{ph}} +  S_{\rm{at}}\frac{({\Delta \nu}{})^2}{4(\nu - \nu_L)^2 +  ({\Delta \nu}{})^2 }
\label{eq:fitfunction}
\end{equation}
where  $S_{\rm{ph}}$ is the shot-noise background, $S_{\rm{at}}$ is the peak atomic noise contribution, $\nu_L$ is the $^{85}$Rb Larmor frequency,  and $\Delta \nu$ is the magnetic resonance linewidth, thus $ {\bf v} \equiv (S_{\rm{ph}},\nu_L,S_{\rm{at}},\Delta \nu)$. As described in the SI and in  \cite{Lucivero2016}, ${\bf v}$ can be independently computed from the atomic number density $n$ and probe power $P$, allowing first-principles computation of the covariance matrix $\GamTh$ through Eq.~(\ref{eq:Sensitivity}). We define the \textit{spin noise signal-to-noise ratio} as ${\rm SNR} \equiv S_{\rm at}/S_{\rm ph}$.
%$S_{\rm{ph}}$ and $\Delta \nu$ depend on the probe power $P$, and $\Delta \nu$ depends also on atomic density $n$.
%A full description of the atomic and optical physics is given in \cite{Lucivero2016}.

A representative experimental spectrum, with $n = \unit{4.23\times 10^{12}}{\centi\meter^{-3}}$, $P = \unit{2.0}{\milli\watt}$, $T = \unit{500}{\milli\second}$ and $\Delta = \unit{5}{\micro\second}$, is shown in Fig. \ref{fig:Setup}, after averaging $\Nave = 5$ and coarse-graining $\Nbin = 20$.  In these same conditions we acquired $N = 100$ such spin noise spectra with $\Nave = 1$ and $\Nbin = 50$, and performed maximum-likelihood fits of $f(\nu,{\bf v})$ from Eq.~(\ref{eq:fitfunction}) over the range $33-52$ kHz, which covers the $^{85}$Rb resonance at $\nu_L=42.6$ kHz.  This gives  $N=100$
%\ctext{I think this N here should be used to represent N = $\Nave\Nbin$, not not N= $\Nave$  MWM: it is not either of those things. ${\cal N} = \Nave\Nbin$ indicates the amount of averaging.  Here $N$ is the size of the sample.  Need to say that more clearly}
samples ${\bf v}^{(i)}$ %, i = 1, \ldots, N$
of the vector $\mathbf{v}$, from which we calculate the sample mean $\bar{\bf v} \equiv \sum_{i=1}^{N} {\bf v}^{(i)} / N$ and $\GamExp_{ab} \equiv \sum_{i=1}^{N} (v_a^{(i)} - \bar{v}_a^{(i)})(v_b^{(i)} - \bar{v}_b^{(i)}) / N$, the maximum likelihood estimate (MLE) of the covariance matrix. Throughout, $\bar{\bf v}$ is found to follow closely the \textit{a priori} theory given in the SI. The covariance matrix is
\begin{equation}
\GamExp=
\left(
\begin{array}{cccc}
32.7 & 29 & 0.06 & -349 \\
29 & 2290 & 490 & -142000 \\
0.06 & 490 & 3590 & -6680 \\
-349 & -142000 & -6680 & 21500 \\
%32.6724 & 29.3928 & 0.061848 & -348.594 \\
%29.3928 & 2285.93 & 489.837 & -141651. \\
%0.061848 & 489.837 & 3591.09 & -6677.27 \\
%-348.594 & -141651. & -6677.27 & 21484.9 \\
\end{array}
\right)
\label{ExpGamma}
\end{equation}
%\begin{equation} % 105 C results
%\GamExp=
%\left(
%\begin{array}{cccc}
%38.86 & 37.24 & 0.48 & -432.21 \\
%37.24 & 2043.8 & 658.87 & -118595 \\
%0.48 & 658.9 & 3191.8 & -4939.77 \\
%-432.21 & -118595 & -4939.77 & 17051.9 \\
%\end{array}
%\right)
%\label{ExpGamma}
%\end{equation}
where, due to the definition of $\mathbf{v}$, the units are \unit{}{\micro\volt\squared\per\hertz} for  $S_{\rm{ph}}$ and $S_{\rm{at}}$ and \unit{}{\hertz} for $\nu_L$ and $\Delta \nu$, so that, e.g., the $(1,2)$ entry has units $\unit{}{\micro\volt\squared}$. The covariance matrix predicted by Eq.~(\ref{eq:Sensitivity}) is
\begin{equation}
\GamTh=
\left(
\begin{array}{cccc}
32.4 & 0.13 & 0.44 & -387 \\
0.13 & 1990 & 0.11 & -130 \\
0.44 & 0.11 & 3510 & -5790 \\
-387 & -130 & -5790 & 17000 \\
%32.361 & 0.127692 & 0.441652 & -386.771 \\
%0.127692 & 1994.57 & 0.108235 & -126.876 \\
%0.441652 & 0.108235 & 3509.73 & -5791.49 \\
%-386.771 & -126.876 & -5791.49 & 16974.8 \\
\end{array}
\right).
\label{TheoGamma}
\end{equation}

%\begin{equation}  % 105 C results
%\GamTh=
%\left(
%\begin{array}{cccc}
%38.61 & 0.21 & 0.55 & -452.42 \\
%0.21 & 1981.9 & 0.18 & -199.1 \\
%0.55 & 0.18 & 3839.5 & -6137.14 \\
%-452.42 & -199.1 & -6137.14 & 17409.7 \\
%\end{array}
%\right).
%\label{TheoGamma}
%\end{equation}

To compare these, we note that $N$ %\ctext{$n$ is already being used to denote density above, maybe change to $m$?}
$p$-dimensional vectors drawn from a multivariate normal distribution with covariance matrix $\GamTh$ give rise to a MLE covariance matrix $\GamMLE$ with ${N}\GamMLE$ described by the Wishart distribution $W_p(\GamTh,N)$,
%(\ctext{maybe $m$ instead of $n$?: MmW changed $n$ to $N$, which is already in use for the number of samples}})
so that elements of $\GamMLE$ have variances $(\sigma^{\rm th}_{ij})^2 \equiv \var(\GamMLE_{ij}) = [(\GamTh_{ij})^2 + \GamTh_{ii}\GamTh_{jj}]/N$ \cite{Wishart1928}. Explicitly, the standard deviations are
\begin{equation}
\sigma^{\rm th}=
%\left(
%\begin{array}{cccc}
%4.57654 & 25.406 & 0.339895 & 83.601 \\
%25.406 & 282.075 & 264.583 & 58187.1 \\
%0.339895 & 264.583 & 496.35 & 964.978 \\
%83.601 & 58187.1 & 964.978 & 2400.6 \\
%\end{array}
%\right)
\left(
\begin{array}{cccc}
4.6 & 25 & 0.34 & 84 \\
25 & 280 & 260 & 58000 \\
0.34 & 260 & 500 & 960 \\
84 & 58000 & 960 & 2400 \\
\end{array}
\right)
\label{StdGamma}
\end{equation}
%\begin{equation} % 105 C result
%\sigma^{\rm th}=
%\left(
%\begin{array}{cccc}
%5.18 & 25.88 & 0.39 & 86.35 \\
%25.88 & 258.74 & 275.66 & 53653.6 \\
%0.39 & 275.66 & 587.39 & 1013.94 \\
%86.35 & 53653.6 & 1013.94 & 2225.18 \\
%\end{array}
%\right).
%\label{StdGamma}
%\end{equation}
with the same units as $\GamTh$, $\GamExp$. The normalized absolute deviation $|\GamTh_{ij}-\Gamma_{ij}^{\rm exp}|/ \sigma_{ij}^{\rm th}$ is $\sim 1$ for all elements, with the largest such deviation being $2.4$, indicating good agreement of theory with experiment.  Acquiring and analyzing spectra over the $(n,P)$ ranges $1.49\times 10^{12}$ cm$^{-3} \le n \le 12.6\times 10^{12}$ cm$^{-3}$ and $\unit{0.5}{\milli\watt} \le P \le 4$ mW,  we find that similar agreement is seen over the whole range. This is the second main result of the work: the experimental validation of Eq. (\ref{eq:Sensitivity}).

%\left(
%\begin{array}{cccc}
%0.0680369 & 1.1519 & -1.11741 & 0.456656 \\
%1.1519 & 1.0329 & 1.85095 & -2.43222 \\
%-1.11741 & 1.85095 & 0.163915 & -0.917926 \\
%0.456656 & -2.43222 & -0.917926 & 1.87876 \\
%\end{array}
%\right)

\begin{figure}[t]
\centering
\includegraphics[width=\columnwidth]{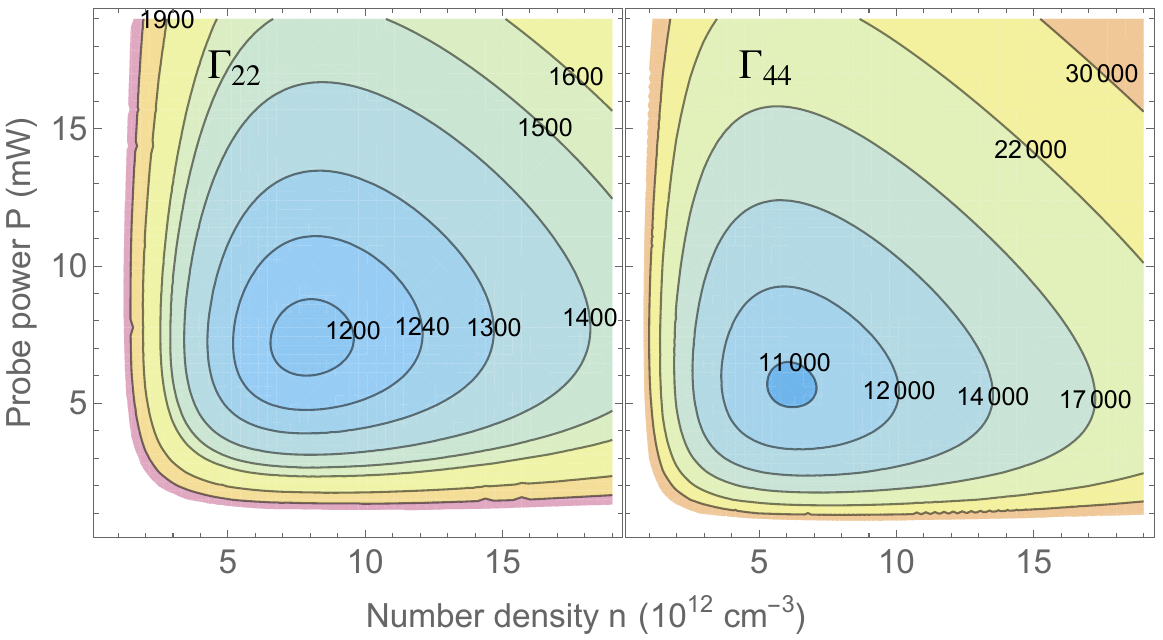}
  \caption{ $\GamTh_{22}$ (left), variance in estimated line center and $\GamTh_{44}$ (right), variance in estimated linewidth  as predicted by Eq.~(\ref{eq:Sensitivity}) as a function of number density and probe power.  Both in units of \unit{}{\hertz\squared}.  The respective optima are $\GamTh_{22}= \unit{1190}{\hertz\squared}$ and $\GamTh_{44}= \unit{10914}{\hertz\squared}$.
}
  \label{Fig:ContourPlots}
\end{figure}

\begin{figure}[t]
\centering
  \hspace{0.4cm}
  {\includegraphics[width=0.9 \columnwidth]{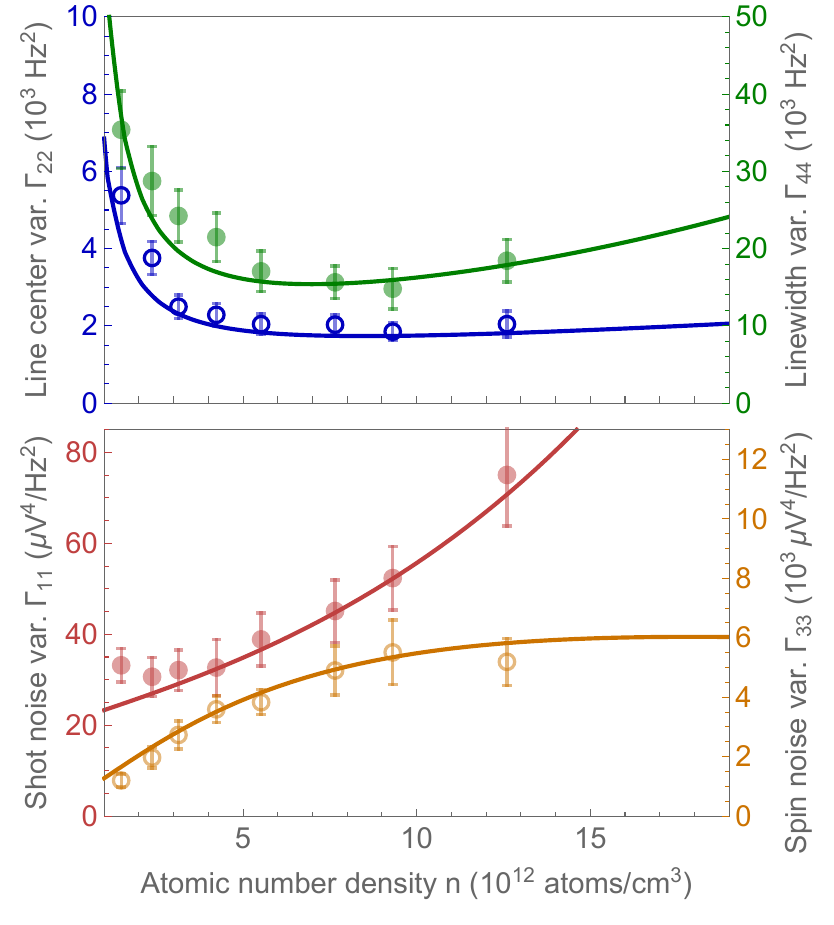}}
  \caption{Sensitivity of spin noise spectroscopy versus atomic density in theory and experiment. Optical power is $P=2$ mW throughout.  (a) Lower, blue curve shows $\Gamma_{22}$, the variance of the Larmor frequency estimate, computed by Eq.~(\ref{eq:Sensitivity}) and from experiment (blue hollow circles), on left (blue) axis.    %The same is shown for $\Gamma_{11}$, the variance of the optical noise power estimate (inset).
  Upper, green curve shows  $\Gamma_{44}$, the variance of the linewidth estimate, and observed variance (green filled circles), on right (green) axis.  (b) Upper, red curve shows $\Gamma_{11}$, the variance of the shot noise estimate, from theory and from experiment (red filled circles), on left (red) axis.
  %The same is shown for $\Gamma_{11}$, the variance of the optical noise power estimate (inset).
  Lower, orange curve shows  $\Gamma_{33}$, the variance of the spin noise estimate, and observed variance (orange hollow circles), on right (orange) axis.  Error bars show plus/minus one standard error (see SI).
  % and $\Gamma_{33}$, the variance of the  atomic noise strength  estimate (inset).
% In each plot, black dashed and brown dot-dashed curves show theory with one contribution; atomic or optical, respectively, set ``by hand''  to zero.  This shows the nonlinear relationship between physical noise sources and noise in the fit parameters, and the potential gain from squeezing. \mtext{I'm not sure I understand this setting ``by hand'', and in any case it is not described anywhere in the manuscript.}
%     Inset shows Experimental covariance matrix term $\GamExp_{11}(n)$ (blue filled circles). \textbf{(b)} Experimental covariance matrix term $\GamExp_{44}(n)$ (blue filled circles) versus atomic density. \textbf{Inset (b)} Experimental covariance matrix term $\GamExp_{33}(n)$ (black filled circles). The continuous red lines represent the theoretical bound to the variances (SQL).  The dashed and dot-dashed curves show optical-noise-only and atomic spin-noise-only scenarios, respectively. Error bars show the standard errors of the sample covariance term, calculated by applying the $k$-statistics \cite{KenneyBook1940} to the individual parameter distribution (see text).
  }
  \label{Fig:ExpVarGroup4Plots}
\end{figure}

%This is what GL wrote about the brown and black curves.  I don't see the motivation for the procedure he describes, which is to subtract one noise source *only from the denominator* of the fraction in Eq.~(3).  Removing the optical noise, for example, would also change the numerator, since it would change the noise spectrum itself. I propose to replace these black/brown curves with a single curve calculated with $\xi^2=0$.  That will indicate the ``quantum contribution'' (if the atomic noise source is considered classical) and the limit of what might be achievable through squeezing.
% In all the plots of Fig. (\ref{Fig:ExpVarGroup4Plots}) we show two functions (black dashed and brown dot-dashed) that are obtained by calculating Eq.~(\ref{eq:CovMat}) when either the uncertainty $\sigma(\nu)= \sigma_{\rm ph}$ or $\sigma(\nu)= \sigma_{\rm at}(\nu)$ is inserted into the denominator of Eq.~(\ref{eq:Mjk}), respectively. The resulting functions show qualitatively that the SQL of optically-detected noise spectroscopy arises from both quantum fluctuations of the probe photons (shot-noise) and intrinsic fluctuations of the investigated system, resonant spin-noise in the specific case. Then, we claim that the increasing of the variances after the optimal region, in the high-density shown in Figs. (\ref{Fig:ContourPlots}) and (\ref{Fig:ExpVarGroup4Plots}), is due to the increased atomic noise contribution to the spectrum uncertainty together with additional linewidth broadening due to atomic collisions \cite{Lucivero2016}.

\PRLsection{Applications in spin noise spectroscopy} We now apply {our result for the sensitivity of noise spectroscopy,} Eq.~(\ref{eq:Sensitivity}), to study the standard quantum limits of {FR-SNS}. %\ctext{optically-detected spin noise spectroscopy}.
For any given $(n,P)$, the sensitivity given by Eq.~(\ref{eq:Sensitivity}) and Eq.~(\ref{eq:fitfunction}) describes a standard quantum limit, in the sense that the readout noise $S_{\rm ph}$ is at the shot-noise level, and cannot be reduced by classical techniques. We will refer to this as a ``local'' standard quantum limit, because it applies to a specific point in $(n,P)$ space.

{As described in our previous work \cite{Lucivero2016}} the parameters ${\bf v}$ that appear in Eq.~(\ref{eq:fitfunction}) have the following dependencies on $(n,P)$:  $S_{\rm{at}} \propto P^2 n/\Delta \nu$,  $\Delta \nu = \Delta \nu_0 +A n + B P$, where $A$ and $B$ are constants describing collisional and power broadening, respectively, and $S_{\rm{ph}} \propto P \xi^2$, where $\xi^2$ is the squeezing parameter, with $\xi^2 = 1$ indicating the shot noise level.  $\nu_L$ is to good approximation independent of $(n,P)$ in the ranges we study. This form, a white noise background plus a Lorentzian spectral feature subject to line broadening with increasing probe power is found in solid state \cite{Glasenapp2013} as well as atomic systems.  Both  broadening and narrowing of resonance lines with increasing dopant concentration is observed in solid state systems \cite{Hubner2014}.

These $(n,P)$ dependencies, along with the fundamental sensitivity given in Eq.~(\ref{eq:Sensitivity}), create global optima for the variances of $\nu_L$ and $\Delta \nu$, and thus the sensitivity to these parameters. The optimum can be understood via the dependence of SNR and line broadening on $P$ and $n$: at low values, SNR increases as the product of these, while line broadening is negligible. At high $P$ or $n$, the SNR saturates while the line broadening increases without limit, reducing sensitivity. An intermediate condition gives the optimum. Similar line broadening considerations determine the optima for optical magnetometers \cite{BudkerPRL1998,Budker2007,PustelnyJAP2008}.

As shown in Fig. \ref{Fig:ContourPlots}, for our $^{85}$Rb system, the optima are near $(n,P) =  (\unit{7 \times 10^{12}}{\centi\meter^{-3}}, \unit{7}{\milli\watt})$. Experimental exploration of these optima is shown in  Figs. \ref{Fig:ExpVarGroup4Plots} and  \ref{Fig:SqData}. These represent the standard quantum limit, not for a particular $(n,P)$, but rather over the whole $(n,P)$ parameter space. For this reason, we refer to these as the ``global'' standard quantum limits for our system. This is our third main result, that there exist global standard quantum limits in SNS.

%\rtext{Indeed, in the low (n,P) regime, increasing the atom/photon number reduces the uncertainty in the estimation of $\nu_L$ and $\Delta \nu$ because of the improved SNR \cite{Lucivero2016}. The optima occur when the spin noise signal of Fig. \ref{fig:Setup} starts to saturate in addition to collisional/power linewidth broadening. Further increasing of both power and density just adds uncertainty to the noisy spectrum, further ``noise of the noise'', resulting in an inversion above the optima i.e. reduced sensitivity, as predicted by Eq.~(\ref{eq:Sensitivity}). A similar tradeoff between signal amplitude and resonance linewidth is typical in optimization of optical magnetometers cite{OpticalMagnetometry,NarrowBudker,AMORPustelny}.
%}
%\mtext{explain why the curve has the shape it does.}
%

\begin{figure}[t]
\centering
  \hspace{0.4cm}
  {\includegraphics[width=0.99 \columnwidth]{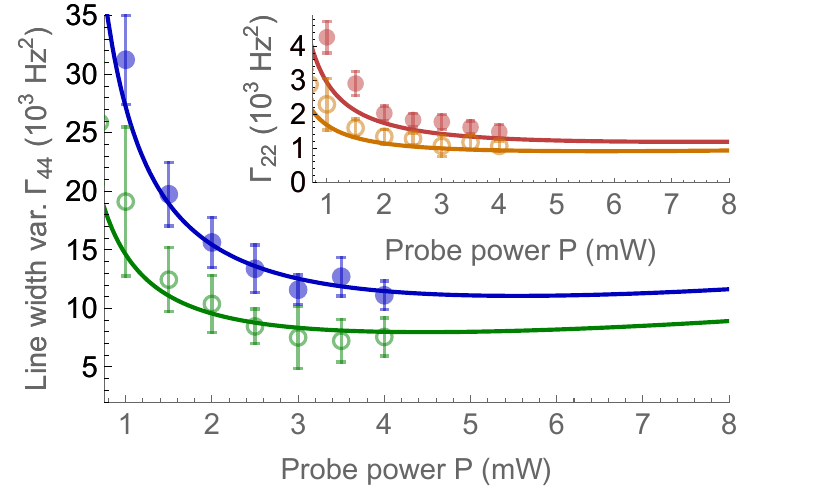}}
%  {\includegraphics[width=0.8 \columnwidth]{../images2/Squeezed&CohData160912.pdf}}
  \caption{Sensitivity of spin noise spectroscopy versus probe power, and sensitivity enhancement by squeezing.  Atom number density is $n=7.65\times10^{12}$ cm$^{-3}$ throughout. Main plot:  variance $\GamExp_{44}$ of the linewidth estimate, versus power for a coherent-state probe by Eq.~(\ref{eq:Sensitivity}) (upper, blue curve) and experiment (blue filled circles).   Same quantities with a polarization squeezed probe with \unit{2.6}{dB} of squeezing
%$\xi^2 = 0.55$
by Eq.~(\ref{eq:Sensitivity}) (green, lower curve) and experiment (green hollow circles). Error bars show plus/minus one standard error.  Inset shows the same quantities for $\GamExp_{22}$, the variance of the Larmor frequency estimate.
%  The theoretical SQL for $\xi^{2}=1$ (coherent probe) and  $\xi^{2}=0.55$ (squeezed probe) is shown with continuous and dashed curves, respectively. \textbf{Inset} Analogue analysis for the covariance term $\GamExp_{22}$(P). The fixed atomic density is $n=7.65\times10^{12}$ cm$^{-3}$. Error bars show the standard errors of the sample covariance term, calculated by applying the $k$-statistics \cite{KenneyBook1940} to the individual parameter distribution (see text).
  }
\label{Fig:SqData}
\end{figure}

\PRLsection{Sensitivity enhancement by squeezing}
The existence of global optima motivates the use of optical squeezing, which promises a way to improve statistical sensitivity when the improvement by choice of particle number is exhausted. It is not a foregone conclusion that squeezing will help, however: in other applications of squeezed light to probe high-density ensembles \cite{HorromPRA2012}, squeezing had no beneficial effect at the optimum.  %In light of the non-linear relationships among particle numbers $(n,P)$, spectral properties $S_{\rm at}$ and $\Delta \nu$, and sensitivities $\GamTh$, it is perhaps not obvious that squeezing, which only reduces $S_{\rm ph}$, will help.
%In Fig. \ref{Fig:SqData}, we show that squeezing is effective in this scenario.
Using a sub-threshold OPO described in \cite{PredojevicPRA2008,Lucivero2016}, we generate \unit{2.6}{dB} of polarization squeezing, or $\xi^2 = 0.55$, and observe a reduction of $\GamExp_{22}$ and $\GamExp_{44}$ by a factor $0.61$, constant to within uncertainties, as shown in Fig. \ref{Fig:SqData}. The four $\GamExp_{44}$ measurements from \unit{2.5}{\milli\watt} to \unit{4}{\milli\watt} are in total $7.0$ standard deviations below \unit{10914}{\hertz\squared}, the global standard quantum limit for our system. The optimum with squeezing also occurs at lower probe power than the optimum without squeezing. This leads to less line broadening and more generally a less invasive measurement.  This is our fourth main result, that both local and global standard quantum limits of SNS can be surpassed. This differs from the enhancement reported in \cite{Lucivero2016}, because the SNR is not an exhaustive figure of merit of the sensitivity to spectral parameters, which is instead derived in Eq. (\ref{eq:Sensitivity}).

\PRLsection{Conclusions}
We have derived the sensitivity of noise spectroscopies from estimation theory, finding simple expressions for the Fisher information matrix in terms of the spectral model. The result enables rigorous use of noise spectra in cell biology \cite{TolicNorrelykkePRL2004,TaylorNJP2013}, molecular biophysics \cite{KawakamiL2004,Berg-SorensenRSI2004}, geophysics \cite{DescherevskyNHESS2003}, space science \cite{MoncuquetGRL2005}, and quantum information processing \cite{BiercukN2009,AlvarezPRL2011,YugePRL2011,BylanderNPhys2011,MedfordPRL2012}.  For optically-probed particulate systems that show line-broadening, e.g. atomic vapors and quasiparticle systems in solid state physics, the theory predicts global sensitivity optima as a function of particle number density and probe power. These global standard quantum limits define the limiting sensitivity of noise spectroscopy given unlimited classical resources. We quantitatively validate the theory and confirm the prediction of global optima for spin noise spectroscopy of a hot atomic vapor probed by optical Faraday rotation.  Using a polarization-squeezed probe beam, we surpass the global standard quantum limit for this system. % \mtext{revisit at end}

This work was supported by European Research Council (ERC) projects AQUMET (280169) and ERIDIAN (713682); European Union QUIC (641122); Ministerio de Econom'a y Competitividad (MINECO) Severo Ochoa programme (SEV-2015-0522) and projects MAQRO (Ref. FIS2015-68039-P), XPLICA  (FIS2014-62181-EXP); Ag\`{e}ncia de Gesti\'{o} d'Ajuts Universitaris i de Recerca (AGAUR) project (2014-SGR-1295);
%European Regional Development Fund (FEDER) (TEC2013-46168-R);
Fundaci\'{o} Privada CELLEX; this  project  has  received  funding  from the European Union's Horizon 2020 research and innovation programme under the Marie Sklodowska-Curie grant agreement  QUTEMAG  (no.  654339)

%\PRLsection{Appendix}
%
%From $\tilde{Y}(\nu_i) \equiv \sum_m \exp[i \nu_i t_m] Y(t_m)$, and stationarity, i.e. invariance of correlation functions under global time translation, we find  $\langle \tilde{Y}(\nu_i) \rangle = 0$ for all $\nu_i \ne 0$. Moreover, writing $\langle \tilde{Y}^*(\nu_i) \tilde{Y}(\nu_j) \rangle = \sum_{l,m}  \exp[-i \nu_i t_l] \exp[i \nu_j t_m] \langle Y(t_l) Y(t_m) \rangle = \sum_{l,m}  \exp[i (\nu_j - \nu_i) t_l] \exp[i \nu_j \delta t] \langle Y(t_l) Y(t_l + \delta t) \rangle$, which vanishes for $\nu_j \ne \nu_i$.

\bibliographystyle{apsrev4-1}
\bibliography{../SQLofSNS}

%merlin.mbs apsrev4-1.bst 2010-07-25 4.21a (PWD, AO, DPC) hacked
%Control: key (0)
%Control: author (72) initials jnrlst
%Control: editor formatted (1) identically to author
%Control: production of article title (-1) disabled
%Control: page (0) single
%Control: year (1) truncated
%Control: production of eprint (0) enabled
\begin{thebibliography}{52}%
\makeatletter
\providecommand \@ifxundefined [1]{%
 \@ifx{#1\undefined}
}%
\providecommand \@ifnum [1]{%
 \ifnum #1\expandafter \@firstoftwo
 \else \expandafter \@secondoftwo
 \fi
}%
\providecommand \@ifx [1]{%
 \ifx #1\expandafter \@firstoftwo
 \else \expandafter \@secondoftwo
 \fi
}%
\providecommand \natexlab [1]{#1}%
\providecommand \enquote  [1]{``#1''}%
\providecommand \bibnamefont  [1]{#1}%
\providecommand \bibfnamefont [1]{#1}%
\providecommand \citenamefont [1]{#1}%
\providecommand \href@noop [0]{\@secondoftwo}%
\providecommand \href [0]{\begingroup \@sanitize@url \@href}%
\providecommand \@href[1]{\@@startlink{#1}\@@href}%
\providecommand \@@href[1]{\endgroup#1\@@endlink}%
\providecommand \@sanitize@url [0]{\catcode `\\12\catcode `\$12\catcode
  `\&12\catcode `\#12\catcode `\^12\catcode `\_12\catcode `\%12\relax}%
\providecommand \@@startlink[1]{}%
\providecommand \@@endlink[0]{}%
\providecommand \url  [0]{\begingroup\@sanitize@url \@url }%
\providecommand \@url [1]{\endgroup\@href {#1}{\urlprefix }}%
\providecommand \urlprefix  [0]{URL }%
\providecommand \Eprint [0]{\href }%
\providecommand \doibase [0]{http://dx.doi.org/}%
\providecommand \selectlanguage [0]{\@gobble}%
\providecommand \bibinfo  [0]{\@secondoftwo}%
\providecommand \bibfield  [0]{\@secondoftwo}%
\providecommand \translation [1]{[#1]}%
\providecommand \BibitemOpen [0]{}%
\providecommand \bibitemStop [0]{}%
\providecommand \bibitemNoStop [0]{.\EOS\space}%
\providecommand \EOS [0]{\spacefactor3000\relax}%
\providecommand \BibitemShut  [1]{\csname bibitem#1\endcsname}%
\let\auto@bib@innerbib\@empty
%</preamble>
\bibitem [{\citenamefont {Crooker}\ \emph {et~al.}(2004)\citenamefont
  {Crooker}, \citenamefont {Rickel}, \citenamefont {Balatsky},\ and\
  \citenamefont {Smith}}]{Crooker2004}%
  \BibitemOpen
  \bibfield  {author} {\bibinfo {author} {\bibfnamefont {S.~A.}\ \bibnamefont
  {Crooker}}, \bibinfo {author} {\bibfnamefont {D.~G.}\ \bibnamefont {Rickel}},
  \bibinfo {author} {\bibfnamefont {A.~V.}\ \bibnamefont {Balatsky}}, \ and\
  \bibinfo {author} {\bibfnamefont {D.~L.}\ \bibnamefont {Smith}},\ }\href
  {http://dx.doi.org/10.1038/nature02804} {\bibfield  {journal} {\bibinfo
  {journal} {Nature}\ }\textbf {\bibinfo {volume} {431}},\ \bibinfo {pages}
  {49} (\bibinfo {year} {2004})}\BibitemShut {NoStop}%
\bibitem [{\citenamefont {Sikula}\ and\ \citenamefont
  {Stourac}(2002)}]{SikulaMIEL2002}%
  \BibitemOpen
  \bibfield  {author} {\bibinfo {author} {\bibfnamefont {J.}~\bibnamefont
  {Sikula}}\ and\ \bibinfo {author} {\bibfnamefont {L.}~\bibnamefont
  {Stourac}},\ }in\ \href {\doibase 10.1109/MIEL.2002.1003370} {\emph {\bibinfo
  {booktitle} {MIEL 2002. 23rd International Conference on
  Microelectronics}}},\ Vol.~\bibinfo {volume} {2}\ (\bibinfo {year} {2002})\
  pp.\ \bibinfo {pages} {767--772 vol.2}\BibitemShut {NoStop}%
\bibitem [{\citenamefont {Vitusevich}\ \emph {et~al.}(2009)\citenamefont
  {Vitusevich}, \citenamefont {Petrychuk}, \citenamefont {Kurakin},
  \citenamefont {Danylyuk}, \citenamefont {Mayer}, \citenamefont {Bougrioua},
  \citenamefont {Naumov}, \citenamefont {Belyaev},\ and\ \citenamefont
  {Klein}}]{VitusevichJSM2009}%
  \BibitemOpen
  \bibfield  {author} {\bibinfo {author} {\bibfnamefont {S.~A.}\ \bibnamefont
  {Vitusevich}}, \bibinfo {author} {\bibfnamefont {M.~V.}\ \bibnamefont
  {Petrychuk}}, \bibinfo {author} {\bibfnamefont {A.~M.}\ \bibnamefont
  {Kurakin}}, \bibinfo {author} {\bibfnamefont {S.~V.}\ \bibnamefont
  {Danylyuk}}, \bibinfo {author} {\bibfnamefont {D.}~\bibnamefont {Mayer}},
  \bibinfo {author} {\bibfnamefont {Z.}~\bibnamefont {Bougrioua}}, \bibinfo
  {author} {\bibfnamefont {A.~V.}\ \bibnamefont {Naumov}}, \bibinfo {author}
  {\bibfnamefont {A.~E.}\ \bibnamefont {Belyaev}}, \ and\ \bibinfo {author}
  {\bibfnamefont {N.}~\bibnamefont {Klein}},\ }\href
  {http://stacks.iop.org/1742-5468/2009/i=01/a=P01046} {\bibfield  {journal}
  {\bibinfo  {journal} {Journal of Statistical Mechanics: Theory and
  Experiment}\ }\textbf {\bibinfo {volume} {2009}},\ \bibinfo {pages} {P01046}
  (\bibinfo {year} {2009})}\BibitemShut {NoStop}%
\bibitem [{\citenamefont {Romach}\ \emph {et~al.}(2015)\citenamefont {Romach},
  \citenamefont {M\"uller}, \citenamefont {Unden}, \citenamefont {Rogers},
  \citenamefont {Isoda}, \citenamefont {Itoh}, \citenamefont {Markham},
  \citenamefont {Stacey}, \citenamefont {Meijer}, \citenamefont {Pezzagna},
  \citenamefont {Naydenov}, \citenamefont {McGuinness}, \citenamefont
  {Bar-Gill},\ and\ \citenamefont {Jelezko}}]{RomachPRL2015}%
  \BibitemOpen
  \bibfield  {author} {\bibinfo {author} {\bibfnamefont {Y.}~\bibnamefont
  {Romach}}, \bibinfo {author} {\bibfnamefont {C.}~\bibnamefont {M\"uller}},
  \bibinfo {author} {\bibfnamefont {T.}~\bibnamefont {Unden}}, \bibinfo
  {author} {\bibfnamefont {L.~J.}\ \bibnamefont {Rogers}}, \bibinfo {author}
  {\bibfnamefont {T.}~\bibnamefont {Isoda}}, \bibinfo {author} {\bibfnamefont
  {K.~M.}\ \bibnamefont {Itoh}}, \bibinfo {author} {\bibfnamefont
  {M.}~\bibnamefont {Markham}}, \bibinfo {author} {\bibfnamefont
  {A.}~\bibnamefont {Stacey}}, \bibinfo {author} {\bibfnamefont
  {J.}~\bibnamefont {Meijer}}, \bibinfo {author} {\bibfnamefont
  {S.}~\bibnamefont {Pezzagna}}, \bibinfo {author} {\bibfnamefont
  {B.}~\bibnamefont {Naydenov}}, \bibinfo {author} {\bibfnamefont {L.~P.}\
  \bibnamefont {McGuinness}}, \bibinfo {author} {\bibfnamefont
  {N.}~\bibnamefont {Bar-Gill}}, \ and\ \bibinfo {author} {\bibfnamefont
  {F.}~\bibnamefont {Jelezko}},\ }\href {\doibase
  10.1103/PhysRevLett.114.017601} {\bibfield  {journal} {\bibinfo  {journal}
  {Phys. Rev. Lett.}\ }\textbf {\bibinfo {volume} {114}},\ \bibinfo {pages}
  {017601} (\bibinfo {year} {2015})}\BibitemShut {NoStop}%
\bibitem [{\citenamefont {Toli\'{c}-N\o{}rrelykke}\ \emph
  {et~al.}(2004)\citenamefont {Toli\'{c}-N\o{}rrelykke}, \citenamefont
  {Munteanu}, \citenamefont {Thon}, \citenamefont {Oddershede},\ and\
  \citenamefont {Berg-S\o{}rensen}}]{TolicNorrelykkePRL2004}%
  \BibitemOpen
  \bibfield  {author} {\bibinfo {author} {\bibfnamefont {I.~M.}\ \bibnamefont
  {Toli\'{c}-N\o{}rrelykke}}, \bibinfo {author} {\bibfnamefont {E.-L.}\
  \bibnamefont {Munteanu}}, \bibinfo {author} {\bibfnamefont {G.}~\bibnamefont
  {Thon}}, \bibinfo {author} {\bibfnamefont {L.}~\bibnamefont {Oddershede}}, \
  and\ \bibinfo {author} {\bibfnamefont {K.}~\bibnamefont {Berg-S\o{}rensen}},\
  }\href {\doibase 10.1103/PhysRevLett.93.078102} {\bibfield  {journal}
  {\bibinfo  {journal} {Phys. Rev. Lett.}\ }\textbf {\bibinfo {volume} {93}},\
  \bibinfo {pages} {078102} (\bibinfo {year} {2004})}\BibitemShut {NoStop}%
\bibitem [{\citenamefont {Taylor}\ \emph {et~al.}(2013)\citenamefont {Taylor},
  \citenamefont {Knittel},\ and\ \citenamefont {Bowen}}]{TaylorNJP2013}%
  \BibitemOpen
  \bibfield  {author} {\bibinfo {author} {\bibfnamefont {M.~A.}\ \bibnamefont
  {Taylor}}, \bibinfo {author} {\bibfnamefont {J.}~\bibnamefont {Knittel}}, \
  and\ \bibinfo {author} {\bibfnamefont {W.~P.}\ \bibnamefont {Bowen}},\ }\href
  {http://stacks.iop.org/1367-2630/15/i=2/a=023018} {\bibfield  {journal}
  {\bibinfo  {journal} {New Journal of Physics}\ }\textbf {\bibinfo {volume}
  {15}},\ \bibinfo {pages} {023018} (\bibinfo {year} {2013})}\BibitemShut
  {NoStop}%
\bibitem [{\citenamefont {Kawakami}\ \emph {et~al.}(2004)\citenamefont
  {Kawakami}, \citenamefont {Byrne}, \citenamefont {Khatri}, \citenamefont
  {Mcleish}, \citenamefont {Radford},\ and\ \citenamefont
  {Smith}}]{KawakamiL2004}%
  \BibitemOpen
  \bibfield  {author} {\bibinfo {author} {\bibfnamefont {M.}~\bibnamefont
  {Kawakami}}, \bibinfo {author} {\bibfnamefont {K.}~\bibnamefont {Byrne}},
  \bibinfo {author} {\bibfnamefont {B.}~\bibnamefont {Khatri}}, \bibinfo
  {author} {\bibfnamefont {T.~C.~B.}\ \bibnamefont {Mcleish}}, \bibinfo
  {author} {\bibfnamefont {S.~E.}\ \bibnamefont {Radford}}, \ and\ \bibinfo
  {author} {\bibfnamefont {D.~A.}\ \bibnamefont {Smith}},\ }\href {\doibase
  10.1021/la0486178} {\bibfield  {journal} {\bibinfo  {journal} {Langmuir}\
  }\textbf {\bibinfo {volume} {20}},\ \bibinfo {pages} {9299} (\bibinfo {year}
  {2004})}\BibitemShut {NoStop}%
\bibitem [{\citenamefont {Berg-S{\o}rensen}\ and\ \citenamefont
  {Flyvbjerg}(2004)}]{Berg-SorensenRSI2004}%
  \BibitemOpen
  \bibfield  {author} {\bibinfo {author} {\bibfnamefont {K.}~\bibnamefont
  {Berg-S{\o}rensen}}\ and\ \bibinfo {author} {\bibfnamefont {H.}~\bibnamefont
  {Flyvbjerg}},\ }\href
  {http://scitation.aip.org/content/aip/journal/rsi/75/3/10.1063/1.1645654}
  {\bibfield  {journal} {\bibinfo  {journal} {Review of Scientific
  Instruments}\ }\textbf {\bibinfo {volume} {75}},\ \bibinfo {pages} {594}
  (\bibinfo {year} {2004})}\BibitemShut {NoStop}%
\bibitem [{\citenamefont {Descherevsky}\ \emph {et~al.}(2003)\citenamefont
  {Descherevsky}, \citenamefont {Lukk}, \citenamefont {Sidorin}, \citenamefont
  {Vstovsky},\ and\ \citenamefont {Timashev}}]{DescherevskyNHESS2003}%
  \BibitemOpen
  \bibfield  {author} {\bibinfo {author} {\bibfnamefont {A.~V.}\ \bibnamefont
  {Descherevsky}}, \bibinfo {author} {\bibfnamefont {A.~A.}\ \bibnamefont
  {Lukk}}, \bibinfo {author} {\bibfnamefont {A.~Y.}\ \bibnamefont {Sidorin}},
  \bibinfo {author} {\bibfnamefont {G.~V.}\ \bibnamefont {Vstovsky}}, \ and\
  \bibinfo {author} {\bibfnamefont {S.~F.}\ \bibnamefont {Timashev}},\ }\href
  {\doibase 10.5194/nhess-3-159-2003} {\bibfield  {journal} {\bibinfo
  {journal} {Natural Hazards and Earth System Science}\ }\textbf {\bibinfo
  {volume} {3}},\ \bibinfo {pages} {159} (\bibinfo {year} {2003})}\BibitemShut
  {NoStop}%
\bibitem [{\citenamefont {Moncuquet}\ \emph {et~al.}(2005)\citenamefont
  {Moncuquet}, \citenamefont {Lecacheux}, \citenamefont {Meyer-Vernet},
  \citenamefont {Cecconi},\ and\ \citenamefont {Kurth}}]{MoncuquetGRL2005}%
  \BibitemOpen
  \bibfield  {author} {\bibinfo {author} {\bibfnamefont {M.}~\bibnamefont
  {Moncuquet}}, \bibinfo {author} {\bibfnamefont {A.}~\bibnamefont
  {Lecacheux}}, \bibinfo {author} {\bibfnamefont {N.}~\bibnamefont
  {Meyer-Vernet}}, \bibinfo {author} {\bibfnamefont {B.}~\bibnamefont
  {Cecconi}}, \ and\ \bibinfo {author} {\bibfnamefont {W.~S.}\ \bibnamefont
  {Kurth}},\ }\href {\doibase 10.1029/2005GL022508} {\bibfield  {journal}
  {\bibinfo  {journal} {Geophysical Research Letters}\ }\textbf {\bibinfo
  {volume} {32}} (\bibinfo {year} {2005}),\ 10.1029/2005GL022508},\ \bibinfo
  {note} {l20S02}\BibitemShut {NoStop}%
\bibitem [{\citenamefont {Safavi-Naeini}\ \emph {et~al.}(2012)\citenamefont
  {Safavi-Naeini}, \citenamefont {Chan}, \citenamefont {Hill}, \citenamefont
  {Alegre}, \citenamefont {Krause},\ and\ \citenamefont
  {Painter}}]{SafaviPRL2012}%
  \BibitemOpen
  \bibfield  {author} {\bibinfo {author} {\bibfnamefont {A.~H.}\ \bibnamefont
  {Safavi-Naeini}}, \bibinfo {author} {\bibfnamefont {J.}~\bibnamefont {Chan}},
  \bibinfo {author} {\bibfnamefont {J.~T.}\ \bibnamefont {Hill}}, \bibinfo
  {author} {\bibfnamefont {T.~P.~M.}\ \bibnamefont {Alegre}}, \bibinfo {author}
  {\bibfnamefont {A.}~\bibnamefont {Krause}}, \ and\ \bibinfo {author}
  {\bibfnamefont {O.}~\bibnamefont {Painter}},\ }\href {\doibase
  10.1103/PhysRevLett.108.033602} {\bibfield  {journal} {\bibinfo  {journal}
  {Phys. Rev. Lett.}\ }\textbf {\bibinfo {volume} {108}},\ \bibinfo {pages}
  {033602} (\bibinfo {year} {2012})}\BibitemShut {NoStop}%
\bibitem [{\citenamefont {Biercuk}\ \emph {et~al.}(2009)\citenamefont
  {Biercuk}, \citenamefont {Uys}, \citenamefont {VanDevender}, \citenamefont
  {Shiga}, \citenamefont {Itano},\ and\ \citenamefont
  {Bollinger}}]{BiercukN2009}%
  \BibitemOpen
  \bibfield  {author} {\bibinfo {author} {\bibfnamefont {M.~J.}\ \bibnamefont
  {Biercuk}}, \bibinfo {author} {\bibfnamefont {H.}~\bibnamefont {Uys}},
  \bibinfo {author} {\bibfnamefont {A.~P.}\ \bibnamefont {VanDevender}},
  \bibinfo {author} {\bibfnamefont {N.}~\bibnamefont {Shiga}}, \bibinfo
  {author} {\bibfnamefont {W.~M.}\ \bibnamefont {Itano}}, \ and\ \bibinfo
  {author} {\bibfnamefont {J.~J.}\ \bibnamefont {Bollinger}},\ }\href
  {http://dx.doi.org/10.1038/nature07951} {\bibfield  {journal} {\bibinfo
  {journal} {Nature}\ }\textbf {\bibinfo {volume} {458}},\ \bibinfo {pages}
  {996} (\bibinfo {year} {2009})}\BibitemShut {NoStop}%
\bibitem [{\citenamefont {\'Alvarez}\ and\ \citenamefont
  {Suter}(2011)}]{AlvarezPRL2011}%
  \BibitemOpen
  \bibfield  {author} {\bibinfo {author} {\bibfnamefont {G.~A.}\ \bibnamefont
  {\'Alvarez}}\ and\ \bibinfo {author} {\bibfnamefont {D.}~\bibnamefont
  {Suter}},\ }\href {\doibase 10.1103/PhysRevLett.107.230501} {\bibfield
  {journal} {\bibinfo  {journal} {Phys. Rev. Lett.}\ }\textbf {\bibinfo
  {volume} {107}},\ \bibinfo {pages} {230501} (\bibinfo {year}
  {2011})}\BibitemShut {NoStop}%
\bibitem [{\citenamefont {Yuge}\ \emph {et~al.}(2011)\citenamefont {Yuge},
  \citenamefont {Sasaki},\ and\ \citenamefont {Hirayama}}]{YugePRL2011}%
  \BibitemOpen
  \bibfield  {author} {\bibinfo {author} {\bibfnamefont {T.}~\bibnamefont
  {Yuge}}, \bibinfo {author} {\bibfnamefont {S.}~\bibnamefont {Sasaki}}, \ and\
  \bibinfo {author} {\bibfnamefont {Y.}~\bibnamefont {Hirayama}},\ }\href
  {\doibase 10.1103/PhysRevLett.107.170504} {\bibfield  {journal} {\bibinfo
  {journal} {Phys. Rev. Lett.}\ }\textbf {\bibinfo {volume} {107}},\ \bibinfo
  {pages} {170504} (\bibinfo {year} {2011})}\BibitemShut {NoStop}%
\bibitem [{\citenamefont {Bylander}\ \emph {et~al.}(2011)\citenamefont
  {Bylander}, \citenamefont {Gustavsson}, \citenamefont {Yan}, \citenamefont
  {Yoshihara}, \citenamefont {Harrabi}, \citenamefont {Fitch}, \citenamefont
  {Cory}, \citenamefont {Nakamura}, \citenamefont {Tsai},\ and\ \citenamefont
  {Oliver}}]{BylanderNPhys2011}%
  \BibitemOpen
  \bibfield  {author} {\bibinfo {author} {\bibfnamefont {J.}~\bibnamefont
  {Bylander}}, \bibinfo {author} {\bibfnamefont {S.}~\bibnamefont
  {Gustavsson}}, \bibinfo {author} {\bibfnamefont {F.}~\bibnamefont {Yan}},
  \bibinfo {author} {\bibfnamefont {F.}~\bibnamefont {Yoshihara}}, \bibinfo
  {author} {\bibfnamefont {K.}~\bibnamefont {Harrabi}}, \bibinfo {author}
  {\bibfnamefont {G.}~\bibnamefont {Fitch}}, \bibinfo {author} {\bibfnamefont
  {D.~G.}\ \bibnamefont {Cory}}, \bibinfo {author} {\bibfnamefont
  {Y.}~\bibnamefont {Nakamura}}, \bibinfo {author} {\bibfnamefont {J.-S.}\
  \bibnamefont {Tsai}}, \ and\ \bibinfo {author} {\bibfnamefont {W.~D.}\
  \bibnamefont {Oliver}},\ }\href {http://dx.doi.org/10.1038/nphys1994}
  {\bibfield  {journal} {\bibinfo  {journal} {Nat Phys}\ }\textbf {\bibinfo
  {volume} {7}},\ \bibinfo {pages} {565} (\bibinfo {year} {2011})}\BibitemShut
  {NoStop}%
\bibitem [{\citenamefont {Medford}\ \emph {et~al.}(2012)\citenamefont
  {Medford}, \citenamefont {Cywi\ifmmode~\acute{n}\else \'{n}\fi{}ski},
  \citenamefont {Barthel}, \citenamefont {Marcus}, \citenamefont {Hanson},\
  and\ \citenamefont {Gossard}}]{MedfordPRL2012}%
  \BibitemOpen
  \bibfield  {author} {\bibinfo {author} {\bibfnamefont {J.}~\bibnamefont
  {Medford}}, \bibinfo {author} {\bibfnamefont {L.}~\bibnamefont
  {Cywi\ifmmode~\acute{n}\else \'{n}\fi{}ski}}, \bibinfo {author}
  {\bibfnamefont {C.}~\bibnamefont {Barthel}}, \bibinfo {author} {\bibfnamefont
  {C.~M.}\ \bibnamefont {Marcus}}, \bibinfo {author} {\bibfnamefont {M.~P.}\
  \bibnamefont {Hanson}}, \ and\ \bibinfo {author} {\bibfnamefont {A.~C.}\
  \bibnamefont {Gossard}},\ }\href {\doibase 10.1103/PhysRevLett.108.086802}
  {\bibfield  {journal} {\bibinfo  {journal} {Phys. Rev. Lett.}\ }\textbf
  {\bibinfo {volume} {108}},\ \bibinfo {pages} {086802} (\bibinfo {year}
  {2012})}\BibitemShut {NoStop}%
\bibitem [{\citenamefont {Katsoprinakis}\ \emph {et~al.}(2007)\citenamefont
  {Katsoprinakis}, \citenamefont {Dellis},\ and\ \citenamefont
  {Kominis}}]{KatsoprinakisPRA2007}%
  \BibitemOpen
  \bibfield  {author} {\bibinfo {author} {\bibfnamefont {G.~E.}\ \bibnamefont
  {Katsoprinakis}}, \bibinfo {author} {\bibfnamefont {A.~T.}\ \bibnamefont
  {Dellis}}, \ and\ \bibinfo {author} {\bibfnamefont {I.~K.}\ \bibnamefont
  {Kominis}},\ }\href {\doibase 10.1103/PhysRevA.75.042502} {\bibfield
  {journal} {\bibinfo  {journal} {Phys. Rev. A}\ }\textbf {\bibinfo {volume}
  {75}},\ \bibinfo {pages} {042502} (\bibinfo {year} {2007})}\BibitemShut
  {NoStop}%
\bibitem [{\citenamefont {Zapasskii}\ \emph {et~al.}(2013)\citenamefont
  {Zapasskii}, \citenamefont {Greilich}, \citenamefont {Crooker}, \citenamefont
  {Li}, \citenamefont {Kozlov}, \citenamefont {Yakovlev}, \citenamefont
  {Reuter}, \citenamefont {Wieck},\ and\ \citenamefont
  {Bayer}}]{Zapasskii2013a}%
  \BibitemOpen
  \bibfield  {author} {\bibinfo {author} {\bibfnamefont {V.~S.}\ \bibnamefont
  {Zapasskii}}, \bibinfo {author} {\bibfnamefont {A.}~\bibnamefont {Greilich}},
  \bibinfo {author} {\bibfnamefont {S.~A.}\ \bibnamefont {Crooker}}, \bibinfo
  {author} {\bibfnamefont {Y.}~\bibnamefont {Li}}, \bibinfo {author}
  {\bibfnamefont {G.~G.}\ \bibnamefont {Kozlov}}, \bibinfo {author}
  {\bibfnamefont {D.~R.}\ \bibnamefont {Yakovlev}}, \bibinfo {author}
  {\bibfnamefont {D.}~\bibnamefont {Reuter}}, \bibinfo {author} {\bibfnamefont
  {A.~D.}\ \bibnamefont {Wieck}}, \ and\ \bibinfo {author} {\bibfnamefont
  {M.}~\bibnamefont {Bayer}},\ }\href {\doibase 10.1103/PhysRevLett.110.176601}
  {\bibfield  {journal} {\bibinfo  {journal} {Phys. Rev. Lett.}\ }\textbf
  {\bibinfo {volume} {110}},\ \bibinfo {pages} {176601} (\bibinfo {year}
  {2013})}\BibitemShut {NoStop}%
\bibitem [{\citenamefont {Rugar}\ \emph {et~al.}(2004)\citenamefont {Rugar},
  \citenamefont {Budakian}, \citenamefont {Mamin},\ and\ \citenamefont
  {Chui}}]{Rugar2004}%
  \BibitemOpen
  \bibfield  {author} {\bibinfo {author} {\bibfnamefont {D.}~\bibnamefont
  {Rugar}}, \bibinfo {author} {\bibfnamefont {R.}~\bibnamefont {Budakian}},
  \bibinfo {author} {\bibfnamefont {H.~J.}\ \bibnamefont {Mamin}}, \ and\
  \bibinfo {author} {\bibfnamefont {B.~W.}\ \bibnamefont {Chui}},\ }\href
  {http://dx.doi.org/10.1038/nature02658} {\bibfield  {journal} {\bibinfo
  {journal} {Nature}\ }\textbf {\bibinfo {volume} {430}},\ \bibinfo {pages}
  {329} (\bibinfo {year} {2004})}\BibitemShut {NoStop}%
\bibitem [{\citenamefont {Budakian}\ \emph {et~al.}(2005)\citenamefont
  {Budakian}, \citenamefont {Mamin}, \citenamefont {Chui},\ and\ \citenamefont
  {Rugar}}]{Budakian2005}%
  \BibitemOpen
  \bibfield  {author} {\bibinfo {author} {\bibfnamefont {R.}~\bibnamefont
  {Budakian}}, \bibinfo {author} {\bibfnamefont {H.~J.}\ \bibnamefont {Mamin}},
  \bibinfo {author} {\bibfnamefont {B.~W.}\ \bibnamefont {Chui}}, \ and\
  \bibinfo {author} {\bibfnamefont {D.}~\bibnamefont {Rugar}},\ }\href
  {\doibase 10.1126/science.1106718} {\bibfield  {journal} {\bibinfo  {journal}
  {Science}\ }\textbf {\bibinfo {volume} {307}},\ \bibinfo {pages} {408}
  (\bibinfo {year} {2005})}\BibitemShut {NoStop}%
\bibitem [{\citenamefont {Balasubramanian}\ \emph {et~al.}(2008)\citenamefont
  {Balasubramanian}, \citenamefont {Chan}, \citenamefont {Kolesov},
  \citenamefont {Al-Hmoud}, \citenamefont {Tisler}, \citenamefont {Shin},
  \citenamefont {Kim}, \citenamefont {Wojcik}, \citenamefont {Hemmer},
  \citenamefont {Krueger}, \citenamefont {Hanke}, \citenamefont
  {Leitenstorfer}, \citenamefont {Bratschitsch}, \citenamefont {Jelezko},\ and\
  \citenamefont {Wrachtrup}}]{Balasubramanian2008}%
  \BibitemOpen
  \bibfield  {author} {\bibinfo {author} {\bibfnamefont {G.}~\bibnamefont
  {Balasubramanian}}, \bibinfo {author} {\bibfnamefont {I.~Y.}\ \bibnamefont
  {Chan}}, \bibinfo {author} {\bibfnamefont {R.}~\bibnamefont {Kolesov}},
  \bibinfo {author} {\bibfnamefont {M.}~\bibnamefont {Al-Hmoud}}, \bibinfo
  {author} {\bibfnamefont {J.}~\bibnamefont {Tisler}}, \bibinfo {author}
  {\bibfnamefont {C.}~\bibnamefont {Shin}}, \bibinfo {author} {\bibfnamefont
  {C.}~\bibnamefont {Kim}}, \bibinfo {author} {\bibfnamefont {A.}~\bibnamefont
  {Wojcik}}, \bibinfo {author} {\bibfnamefont {P.~R.}\ \bibnamefont {Hemmer}},
  \bibinfo {author} {\bibfnamefont {A.}~\bibnamefont {Krueger}}, \bibinfo
  {author} {\bibfnamefont {T.}~\bibnamefont {Hanke}}, \bibinfo {author}
  {\bibfnamefont {A.}~\bibnamefont {Leitenstorfer}}, \bibinfo {author}
  {\bibfnamefont {R.}~\bibnamefont {Bratschitsch}}, \bibinfo {author}
  {\bibfnamefont {F.}~\bibnamefont {Jelezko}}, \ and\ \bibinfo {author}
  {\bibfnamefont {J.}~\bibnamefont {Wrachtrup}},\ }\href
  {http://dx.doi.org/10.1038/nature07278} {\bibfield  {journal} {\bibinfo
  {journal} {Nature}\ }\textbf {\bibinfo {volume} {455}},\ \bibinfo {pages}
  {648} (\bibinfo {year} {2008})}\BibitemShut {NoStop}%
\bibitem [{\citenamefont {Maze}\ \emph {et~al.}(2008)\citenamefont {Maze},
  \citenamefont {Stanwix}, \citenamefont {Hodges}, \citenamefont {Hong},
  \citenamefont {Taylor}, \citenamefont {Cappellaro}, \citenamefont {Jiang},
  \citenamefont {Dutt}, \citenamefont {Togan}, \citenamefont {Zibrov},
  \citenamefont {Yacoby}, \citenamefont {Walsworth},\ and\ \citenamefont
  {Lukin}}]{Maze2008}%
  \BibitemOpen
  \bibfield  {author} {\bibinfo {author} {\bibfnamefont {J.~R.}\ \bibnamefont
  {Maze}}, \bibinfo {author} {\bibfnamefont {P.~L.}\ \bibnamefont {Stanwix}},
  \bibinfo {author} {\bibfnamefont {J.~S.}\ \bibnamefont {Hodges}}, \bibinfo
  {author} {\bibfnamefont {S.}~\bibnamefont {Hong}}, \bibinfo {author}
  {\bibfnamefont {J.~M.}\ \bibnamefont {Taylor}}, \bibinfo {author}
  {\bibfnamefont {P.}~\bibnamefont {Cappellaro}}, \bibinfo {author}
  {\bibfnamefont {L.}~\bibnamefont {Jiang}}, \bibinfo {author} {\bibfnamefont
  {M.~V.~G.}\ \bibnamefont {Dutt}}, \bibinfo {author} {\bibfnamefont
  {E.}~\bibnamefont {Togan}}, \bibinfo {author} {\bibfnamefont {A.~S.}\
  \bibnamefont {Zibrov}}, \bibinfo {author} {\bibfnamefont {A.}~\bibnamefont
  {Yacoby}}, \bibinfo {author} {\bibfnamefont {R.~L.}\ \bibnamefont
  {Walsworth}}, \ and\ \bibinfo {author} {\bibfnamefont {M.~D.}\ \bibnamefont
  {Lukin}},\ }\href {http://dx.doi.org/10.1038/nature07279} {\bibfield
  {journal} {\bibinfo  {journal} {Nature}\ }\textbf {\bibinfo {volume} {455}},\
  \bibinfo {pages} {644} (\bibinfo {year} {2008})}\BibitemShut {NoStop}%
\bibitem [{\citenamefont {Mamin}\ \emph {et~al.}(2012)\citenamefont {Mamin},
  \citenamefont {Sherwood},\ and\ \citenamefont {Rugar}}]{Mamin2012}%
  \BibitemOpen
  \bibfield  {author} {\bibinfo {author} {\bibfnamefont {H.~J.}\ \bibnamefont
  {Mamin}}, \bibinfo {author} {\bibfnamefont {M.~H.}\ \bibnamefont {Sherwood}},
  \ and\ \bibinfo {author} {\bibfnamefont {D.}~\bibnamefont {Rugar}},\ }\href
  {\doibase 10.1103/PhysRevB.86.195422} {\bibfield  {journal} {\bibinfo
  {journal} {Phys. Rev. B}\ }\textbf {\bibinfo {volume} {86}},\ \bibinfo
  {pages} {195422} (\bibinfo {year} {2012})}\BibitemShut {NoStop}%
\bibitem [{\citenamefont {Aleksandrov}\ and\ \citenamefont
  {Zapasskii}(1981)}]{Zap1981}%
  \BibitemOpen
  \bibfield  {author} {\bibinfo {author} {\bibfnamefont {B.}~\bibnamefont
  {Aleksandrov}}\ and\ \bibinfo {author} {\bibfnamefont {V.~S.}\ \bibnamefont
  {Zapasskii}},\ }\href {http://www.jetp.ac.ru/cgi-bin/dn/e_054_01_0064.pdf}
  {\bibfield  {journal} {\bibinfo  {journal} {Sov. Phys. JETP}\ }\textbf
  {\bibinfo {volume} {54}},\ \bibinfo {pages} {64} (\bibinfo {year}
  {1981})}\BibitemShut {NoStop}%
\bibitem [{\citenamefont {Oestreich}\ \emph {et~al.}(2005)\citenamefont
  {Oestreich}, \citenamefont {R\"omer}, \citenamefont {Haug},\ and\
  \citenamefont {H\"agele}}]{Oestreich2005}%
  \BibitemOpen
  \bibfield  {author} {\bibinfo {author} {\bibfnamefont {M.}~\bibnamefont
  {Oestreich}}, \bibinfo {author} {\bibfnamefont {M.}~\bibnamefont {R\"omer}},
  \bibinfo {author} {\bibfnamefont {R.~J.}\ \bibnamefont {Haug}}, \ and\
  \bibinfo {author} {\bibfnamefont {D.}~\bibnamefont {H\"agele}},\ }\href
  {\doibase 10.1103/PhysRevLett.95.216603} {\bibfield  {journal} {\bibinfo
  {journal} {Phys. Rev. Lett.}\ }\textbf {\bibinfo {volume} {95}},\ \bibinfo
  {pages} {216603} (\bibinfo {year} {2005})}\BibitemShut {NoStop}%
\bibitem [{\citenamefont {Crooker}\ \emph {et~al.}(2010)\citenamefont
  {Crooker}, \citenamefont {Brandt}, \citenamefont {Sandfort}, \citenamefont
  {Greilich}, \citenamefont {Yakovlev}, \citenamefont {Reuter}, \citenamefont
  {Wieck},\ and\ \citenamefont {Bayer}}]{Crooker2010}%
  \BibitemOpen
  \bibfield  {author} {\bibinfo {author} {\bibfnamefont {S.~A.}\ \bibnamefont
  {Crooker}}, \bibinfo {author} {\bibfnamefont {J.}~\bibnamefont {Brandt}},
  \bibinfo {author} {\bibfnamefont {C.}~\bibnamefont {Sandfort}}, \bibinfo
  {author} {\bibfnamefont {A.}~\bibnamefont {Greilich}}, \bibinfo {author}
  {\bibfnamefont {D.~R.}\ \bibnamefont {Yakovlev}}, \bibinfo {author}
  {\bibfnamefont {D.}~\bibnamefont {Reuter}}, \bibinfo {author} {\bibfnamefont
  {A.~D.}\ \bibnamefont {Wieck}}, \ and\ \bibinfo {author} {\bibfnamefont
  {M.}~\bibnamefont {Bayer}},\ }\href {\doibase 10.1103/PhysRevLett.104.036601}
  {\bibfield  {journal} {\bibinfo  {journal} {Phys. Rev. Lett.}\ }\textbf
  {\bibinfo {volume} {104}},\ \bibinfo {pages} {036601} (\bibinfo {year}
  {2010})}\BibitemShut {NoStop}%
\bibitem [{\citenamefont {Glasenapp}\ \emph {et~al.}(2013)\citenamefont
  {Glasenapp}, \citenamefont {Greilich}, \citenamefont {Ryzhov}, \citenamefont
  {Zapasskii}, \citenamefont {Yakovlev}, \citenamefont {Kozlov},\ and\
  \citenamefont {Bayer}}]{Glasenapp2013}%
  \BibitemOpen
  \bibfield  {author} {\bibinfo {author} {\bibfnamefont {P.}~\bibnamefont
  {Glasenapp}}, \bibinfo {author} {\bibfnamefont {A.}~\bibnamefont {Greilich}},
  \bibinfo {author} {\bibfnamefont {I.~I.}\ \bibnamefont {Ryzhov}}, \bibinfo
  {author} {\bibfnamefont {V.~S.}\ \bibnamefont {Zapasskii}}, \bibinfo {author}
  {\bibfnamefont {D.~R.}\ \bibnamefont {Yakovlev}}, \bibinfo {author}
  {\bibfnamefont {G.~G.}\ \bibnamefont {Kozlov}}, \ and\ \bibinfo {author}
  {\bibfnamefont {M.}~\bibnamefont {Bayer}},\ }\href {\doibase
  10.1103/PhysRevB.88.165314} {\bibfield  {journal} {\bibinfo  {journal} {Phys.
  Rev. B}\ }\textbf {\bibinfo {volume} {88}},\ \bibinfo {pages} {165314}
  (\bibinfo {year} {2013})}\BibitemShut {NoStop}%
\bibitem [{\citenamefont {Dellis}\ \emph {et~al.}(2014)\citenamefont {Dellis},
  \citenamefont {Loulakis},\ and\ \citenamefont {Kominis}}]{Dellis2014}%
  \BibitemOpen
  \bibfield  {author} {\bibinfo {author} {\bibfnamefont {A.~T.}\ \bibnamefont
  {Dellis}}, \bibinfo {author} {\bibfnamefont {M.}~\bibnamefont {Loulakis}}, \
  and\ \bibinfo {author} {\bibfnamefont {I.~K.}\ \bibnamefont {Kominis}},\
  }\href {\doibase 10.1103/PhysRevA.90.032705} {\bibfield  {journal} {\bibinfo
  {journal} {Phys. Rev. A}\ }\textbf {\bibinfo {volume} {90}},\ \bibinfo
  {pages} {032705} (\bibinfo {year} {2014})}\BibitemShut {NoStop}%
\bibitem [{\citenamefont {Roy}\ \emph {et~al.}(2015)\citenamefont {Roy},
  \citenamefont {Yang}, \citenamefont {Crooker},\ and\ \citenamefont
  {Sinitsyn}}]{Roy2015}%
  \BibitemOpen
  \bibfield  {author} {\bibinfo {author} {\bibfnamefont {D.}~\bibnamefont
  {Roy}}, \bibinfo {author} {\bibfnamefont {L.}~\bibnamefont {Yang}}, \bibinfo
  {author} {\bibfnamefont {S.~A.}\ \bibnamefont {Crooker}}, \ and\ \bibinfo
  {author} {\bibfnamefont {N.~A.}\ \bibnamefont {Sinitsyn}},\ }\href
  {http://dx.doi.org/10.1038/srep09573} {\bibfield  {journal} {\bibinfo
  {journal} {Scientific Reports}\ }\textbf {\bibinfo {volume} {5}},\ \bibinfo
  {pages} {9573} (\bibinfo {year} {2015})}\BibitemShut {NoStop}%
\bibitem [{\citenamefont {Glasenapp}\ \emph {et~al.}(2014)\citenamefont
  {Glasenapp}, \citenamefont {Sinitsyn}, \citenamefont {Yang}, \citenamefont
  {Rickel}, \citenamefont {Roy}, \citenamefont {Greilich}, \citenamefont
  {Bayer},\ and\ \citenamefont {Crooker}}]{Glasenapp2014}%
  \BibitemOpen
  \bibfield  {author} {\bibinfo {author} {\bibfnamefont {P.}~\bibnamefont
  {Glasenapp}}, \bibinfo {author} {\bibfnamefont {N.}~\bibnamefont {Sinitsyn}},
  \bibinfo {author} {\bibfnamefont {L.}~\bibnamefont {Yang}}, \bibinfo {author}
  {\bibfnamefont {D.}~\bibnamefont {Rickel}}, \bibinfo {author} {\bibfnamefont
  {D.}~\bibnamefont {Roy}}, \bibinfo {author} {\bibfnamefont {A.}~\bibnamefont
  {Greilich}}, \bibinfo {author} {\bibfnamefont {M.}~\bibnamefont {Bayer}}, \
  and\ \bibinfo {author} {\bibfnamefont {S.}~\bibnamefont {Crooker}},\ }\href
  {http://link.aps.org/doi/10.1103/PhysRevLett.113.156601} {\bibfield
  {journal} {\bibinfo  {journal} {Phys. Rev. Lett.}\ }\textbf {\bibinfo
  {volume} {113}},\ \bibinfo {pages} {156601} (\bibinfo {year}
  {2014})}\BibitemShut {NoStop}%
\bibitem [{\citenamefont {Cundiff}\ and\ \citenamefont
  {Mukamel}(2013)}]{Cundiff2013}%
  \BibitemOpen
  \bibfield  {author} {\bibinfo {author} {\bibfnamefont {S.~T.}\ \bibnamefont
  {Cundiff}}\ and\ \bibinfo {author} {\bibfnamefont {S.}~\bibnamefont
  {Mukamel}},\ }\href {\doibase 10.1063/PT.3.2047} {\bibfield  {journal}
  {\bibinfo  {journal} {Physics Today}\ }\textbf {\bibinfo {volume} {66}},\
  \bibinfo {pages} {44} (\bibinfo {year} {2013})}\BibitemShut {NoStop}%
\bibitem [{\citenamefont {Kryvohuz}\ and\ \citenamefont
  {Mukamel}(2015)}]{KryvohuzJCP2015}%
  \BibitemOpen
  \bibfield  {author} {\bibinfo {author} {\bibfnamefont {M.}~\bibnamefont
  {Kryvohuz}}\ and\ \bibinfo {author} {\bibfnamefont {S.}~\bibnamefont
  {Mukamel}},\ }\href {\doibase 10.1063/1.4917527} {\bibfield  {journal}
  {\bibinfo  {journal} {The Journal of Chemical Physics}\ }\textbf {\bibinfo
  {volume} {142}},\ \bibinfo {eid} {212430} (\bibinfo {year} {2015}),\
  10.1063/1.4917527}\BibitemShut {NoStop}%
\bibitem [{\citenamefont {Sinitsyn}\ and\ \citenamefont
  {Pershin}(2016)}]{Sinitsyn2016}%
  \BibitemOpen
  \bibfield  {author} {\bibinfo {author} {\bibfnamefont {N.~A.}\ \bibnamefont
  {Sinitsyn}}\ and\ \bibinfo {author} {\bibfnamefont {Y.~V.}\ \bibnamefont
  {Pershin}},\ }\href {http://stacks.iop.org/0034-4885/79/i=10/a=106501}
  {\bibfield  {journal} {\bibinfo  {journal} {Reports on Progress in Physics}\
  }\textbf {\bibinfo {volume} {79}},\ \bibinfo {pages} {106501} (\bibinfo
  {year} {2016})}\BibitemShut {NoStop}%
\bibitem [{\citenamefont {Lucivero}\ \emph {et~al.}(2016)\citenamefont
  {Lucivero}, \citenamefont {Jim\'enez-Mart\'{\i}nez}, \citenamefont {Kong},\
  and\ \citenamefont {Mitchell}}]{Lucivero2016}%
  \BibitemOpen
  \bibfield  {author} {\bibinfo {author} {\bibfnamefont {V.~G.}\ \bibnamefont
  {Lucivero}}, \bibinfo {author} {\bibfnamefont {R.}~\bibnamefont
  {Jim\'enez-Mart\'{\i}nez}}, \bibinfo {author} {\bibfnamefont
  {J.}~\bibnamefont {Kong}}, \ and\ \bibinfo {author} {\bibfnamefont {M.~W.}\
  \bibnamefont {Mitchell}},\ }\href {\doibase 10.1103/PhysRevA.93.053802}
  {\bibfield  {journal} {\bibinfo  {journal} {Phys. Rev. A}\ }\textbf {\bibinfo
  {volume} {93}},\ \bibinfo {pages} {053802} (\bibinfo {year}
  {2016})}\BibitemShut {NoStop}%
\bibitem [{\citenamefont {Ng}\ \emph {et~al.}(2016)\citenamefont {Ng},
  \citenamefont {Ang}, \citenamefont {Wheatley}, \citenamefont {Yonezawa},
  \citenamefont {Furusawa}, \citenamefont {Huntington},\ and\ \citenamefont
  {Tsang}}]{NgPRA2016}%
  \BibitemOpen
  \bibfield  {author} {\bibinfo {author} {\bibfnamefont {S.}~\bibnamefont
  {Ng}}, \bibinfo {author} {\bibfnamefont {S.~Z.}\ \bibnamefont {Ang}},
  \bibinfo {author} {\bibfnamefont {T.~A.}\ \bibnamefont {Wheatley}}, \bibinfo
  {author} {\bibfnamefont {H.}~\bibnamefont {Yonezawa}}, \bibinfo {author}
  {\bibfnamefont {A.}~\bibnamefont {Furusawa}}, \bibinfo {author}
  {\bibfnamefont {E.~H.}\ \bibnamefont {Huntington}}, \ and\ \bibinfo {author}
  {\bibfnamefont {M.}~\bibnamefont {Tsang}},\ }\href {\doibase
  10.1103/PhysRevA.93.042121} {\bibfield  {journal} {\bibinfo  {journal} {Phys.
  Rev. A}\ }\textbf {\bibinfo {volume} {93}},\ \bibinfo {pages} {042121}
  (\bibinfo {year} {2016})}\BibitemShut {NoStop}%
\bibitem [{\citenamefont {Giovannetti}\ \emph {et~al.}(2004)\citenamefont
  {Giovannetti}, \citenamefont {Lloyd},\ and\ \citenamefont
  {Maccone}}]{GiovannettiS2004}%
  \BibitemOpen
  \bibfield  {author} {\bibinfo {author} {\bibfnamefont {V.}~\bibnamefont
  {Giovannetti}}, \bibinfo {author} {\bibfnamefont {S.}~\bibnamefont {Lloyd}},
  \ and\ \bibinfo {author} {\bibfnamefont {L.}~\bibnamefont {Maccone}},\ }\href
  {\doibase 10.1126/science.1104149} {\bibfield  {journal} {\bibinfo  {journal}
  {Science}\ }\textbf {\bibinfo {volume} {306}},\ \bibinfo {pages} {1330}
  (\bibinfo {year} {2004})}\BibitemShut {NoStop}%
\bibitem [{\citenamefont {Lang}\ and\ \citenamefont
  {Caves}(2013)}]{LangPRL2013}%
  \BibitemOpen
  \bibfield  {author} {\bibinfo {author} {\bibfnamefont {M.~D.}\ \bibnamefont
  {Lang}}\ and\ \bibinfo {author} {\bibfnamefont {C.~M.}\ \bibnamefont
  {Caves}},\ }\href {\doibase 10.1103/PhysRevLett.111.173601} {\bibfield
  {journal} {\bibinfo  {journal} {Phys. Rev. Lett.}\ }\textbf {\bibinfo
  {volume} {111}},\ \bibinfo {pages} {173601} (\bibinfo {year}
  {2013})}\BibitemShut {NoStop}%
\bibitem [{\citenamefont {Predojevic}\ \emph {et~al.}(2008)\citenamefont
  {Predojevic}, \citenamefont {Zhai}, \citenamefont {Caballero},\ and\
  \citenamefont {Mitchell}}]{PredojevicPRA2008}%
  \BibitemOpen
  \bibfield  {author} {\bibinfo {author} {\bibfnamefont {A.}~\bibnamefont
  {Predojevic}}, \bibinfo {author} {\bibfnamefont {Z.}~\bibnamefont {Zhai}},
  \bibinfo {author} {\bibfnamefont {J.~M.}\ \bibnamefont {Caballero}}, \ and\
  \bibinfo {author} {\bibfnamefont {M.~W.}\ \bibnamefont {Mitchell}},\ }\href
  {\doibase 10.1103/PhysRevA.78.063820} {\bibfield  {journal} {\bibinfo
  {journal} {Phys. Rev. A}\ }\textbf {\bibinfo {volume} {78}},\ \bibinfo
  {pages} {063820} (\bibinfo {year} {2008})}\BibitemShut {NoStop}%
\bibitem [{\citenamefont {Wolfgramm}\ \emph {et~al.}(2010)\citenamefont
  {Wolfgramm}, \citenamefont {Cer\`e}, \citenamefont {Beduini}, \citenamefont
  {Predojevi\ifmmode~\acute{c}\else \'{c}\fi{}}, \citenamefont {Koschorreck},\
  and\ \citenamefont {Mitchell}}]{WolfgrammPRL2010}%
  \BibitemOpen
  \bibfield  {author} {\bibinfo {author} {\bibfnamefont {F.}~\bibnamefont
  {Wolfgramm}}, \bibinfo {author} {\bibfnamefont {A.}~\bibnamefont {Cer\`e}},
  \bibinfo {author} {\bibfnamefont {F.~A.}\ \bibnamefont {Beduini}}, \bibinfo
  {author} {\bibfnamefont {A.}~\bibnamefont {Predojevi\ifmmode~\acute{c}\else
  \'{c}\fi{}}}, \bibinfo {author} {\bibfnamefont {M.}~\bibnamefont
  {Koschorreck}}, \ and\ \bibinfo {author} {\bibfnamefont {M.~W.}\ \bibnamefont
  {Mitchell}},\ }\href {\doibase 10.1103/PhysRevLett.105.053601} {\bibfield
  {journal} {\bibinfo  {journal} {Phys. Rev. Lett.}\ }\textbf {\bibinfo
  {volume} {105}},\ \bibinfo {pages} {053601} (\bibinfo {year}
  {2010})}\BibitemShut {NoStop}%
\bibitem [{\citenamefont {M{\"u}ller}\ \emph {et~al.}(2010)\citenamefont
  {M{\"u}ller}, \citenamefont {R{\"o}mer}, \citenamefont {H{\"u}bner},\ and\
  \citenamefont {Oestreich}}]{MullerAPL2010}%
  \BibitemOpen
  \bibfield  {author} {\bibinfo {author} {\bibfnamefont {G.~M.}\ \bibnamefont
  {M{\"u}ller}}, \bibinfo {author} {\bibfnamefont {M.}~\bibnamefont
  {R{\"o}mer}}, \bibinfo {author} {\bibfnamefont {J.}~\bibnamefont
  {H{\"u}bner}}, \ and\ \bibinfo {author} {\bibfnamefont {M.}~\bibnamefont
  {Oestreich}},\ }\bibfield  {booktitle} {\emph {\bibinfo {booktitle} {Applied
  Physics Letters}},\ }\href {\doibase 10.1063/1.3505342} {\bibfield  {journal}
  {\bibinfo  {journal} {Applied Physics Letters}\ }\textbf {\bibinfo {volume}
  {97}},\ \bibinfo {pages} {192109} (\bibinfo {year} {2010})}\BibitemShut
  {NoStop}%
\bibitem [{\citenamefont {Root}\ and\ \citenamefont
  {Pitcher}(1955)}]{RootAMS1955}%
  \BibitemOpen
  \bibfield  {author} {\bibinfo {author} {\bibfnamefont {W.~L.}\ \bibnamefont
  {Root}}\ and\ \bibinfo {author} {\bibfnamefont {T.~S.}\ \bibnamefont
  {Pitcher}},\ }\href {\doibase 10.1214/aoms/1177728548} {\bibfield  {journal}
  {\bibinfo  {journal} {Ann. Math. Statist.}\ }\textbf {\bibinfo {volume}
  {26}},\ \bibinfo {pages} {313} (\bibinfo {year} {1955})}\BibitemShut
  {NoStop}%
\bibitem [{\citenamefont {Hamidi}\ and\ \citenamefont
  {Pearl}(1975)}]{HamidiIEEETIT1975}%
  \BibitemOpen
  \bibfield  {author} {\bibinfo {author} {\bibfnamefont {M.}~\bibnamefont
  {Hamidi}}\ and\ \bibinfo {author} {\bibfnamefont {J.}~\bibnamefont {Pearl}},\
  }\href {\doibase 10.1109/TIT.1975.1055403} {\bibfield  {journal} {\bibinfo
  {journal} {IEEE Transactions on Information Theory}\ }\textbf {\bibinfo
  {volume} {21}},\ \bibinfo {pages} {480} (\bibinfo {year} {1975})}\BibitemShut
  {NoStop}%
\bibitem [{\citenamefont {Levin}(1965)}]{LevinIEEETIT1965}%
  \BibitemOpen
  \bibfield  {author} {\bibinfo {author} {\bibfnamefont {M.}~\bibnamefont
  {Levin}},\ }\href {\doibase 10.1109/TIT.1965.1053714} {\bibfield  {journal}
  {\bibinfo  {journal} {IEEE Transactions on Information Theory}\ }\textbf
  {\bibinfo {volume} {11}},\ \bibinfo {pages} {100} (\bibinfo {year}
  {1965})}\BibitemShut {NoStop}%
\bibitem [{\citenamefont {Klein}\ \emph {et~al.}(2000)\citenamefont {Klein},
  \citenamefont {M\'{e}lard},\ and\ \citenamefont {Zahaf}}]{KleinLAA2000}%
  \BibitemOpen
  \bibfield  {author} {\bibinfo {author} {\bibfnamefont {A.}~\bibnamefont
  {Klein}}, \bibinfo {author} {\bibfnamefont {G.}~\bibnamefont {M\'{e}lard}}, \
  and\ \bibinfo {author} {\bibfnamefont {T.}~\bibnamefont {Zahaf}},\ }\href
  {\doibase 10.1016/S0024-3795(99)00045-2} {\bibfield  {journal} {\bibinfo
  {journal} {Linear Algebra and its Applications}\ }\textbf {\bibinfo {volume}
  {321}},\ \bibinfo {pages} {209 } (\bibinfo {year} {2000})},\ \bibinfo {note}
  {eighth Special Issue on Linear Algebra and Statistics}\BibitemShut {NoStop}%
\bibitem [{Note1()}]{Note1}%
  \BibitemOpen
  \bibinfo {note} {A similar result is given in Eq.~(2.32) of \cite
  {NgPRA2016}. Two important points of difference: The expression in \cite
  {NgPRA2016} describes the \protect \textit {quantum Fisher information},
  i.e., the information available under the best possible measurement, while
  ours indicates the Fisher information \protect \textit {per se}, the
  information of the measurement actually made. Also, our result involves the
  parameter dependence of $f$, including noise due to the probe, whereas that
  of \cite {NgPRA2016} includes only $S_X$, the noise of the external
  force.}\BibitemShut {Stop}%
\bibitem [{\citenamefont {Wishart}(1928)}]{Wishart1928}%
  \BibitemOpen
  \bibfield  {author} {\bibinfo {author} {\bibfnamefont {J.}~\bibnamefont
  {Wishart}},\ }\href {\doibase 10.1093/biomet/20A.1-2.32} {\bibfield
  {journal} {\bibinfo  {journal} {Biometrika}\ }\textbf {\bibinfo {volume} {20A
  (1-2)}},\ \bibinfo {pages} {32} (\bibinfo {year} {1928})}\BibitemShut
  {NoStop}%
\bibitem [{\citenamefont {H\"{u}bner}\ \emph {et~al.}(2014)\citenamefont
  {H\"{u}bner}, \citenamefont {Berski}, \citenamefont {Dahbashi},\ and\
  \citenamefont {Oestreich}}]{Hubner2014}%
  \BibitemOpen
  \bibfield  {author} {\bibinfo {author} {\bibfnamefont {J.}~\bibnamefont
  {H\"{u}bner}}, \bibinfo {author} {\bibfnamefont {F.}~\bibnamefont {Berski}},
  \bibinfo {author} {\bibfnamefont {R.}~\bibnamefont {Dahbashi}}, \ and\
  \bibinfo {author} {\bibfnamefont {M.}~\bibnamefont {Oestreich}},\ }\href
  {\doibase 10.1002/pssb.201350291} {\bibfield  {journal} {\bibinfo  {journal}
  {physica status solidi (b)}\ }\textbf {\bibinfo {volume} {251}},\ \bibinfo
  {pages} {1824} (\bibinfo {year} {2014})}\BibitemShut {NoStop}%
\bibitem [{\citenamefont {Budker}\ \emph {et~al.}(1998)\citenamefont {Budker},
  \citenamefont {Yashchuk},\ and\ \citenamefont {Zolotorev}}]{BudkerPRL1998}%
  \BibitemOpen
  \bibfield  {author} {\bibinfo {author} {\bibfnamefont {D.}~\bibnamefont
  {Budker}}, \bibinfo {author} {\bibfnamefont {V.}~\bibnamefont {Yashchuk}}, \
  and\ \bibinfo {author} {\bibfnamefont {M.}~\bibnamefont {Zolotorev}},\ }\href
  {\doibase 10.1103/PhysRevLett.81.5788} {\bibfield  {journal} {\bibinfo
  {journal} {Phys. Rev. Lett.}\ }\textbf {\bibinfo {volume} {81}},\ \bibinfo
  {pages} {5788} (\bibinfo {year} {1998})}\BibitemShut {NoStop}%
\bibitem [{\citenamefont {Budker}\ and\ \citenamefont
  {Romalis}(2007)}]{Budker2007}%
  \BibitemOpen
  \bibfield  {author} {\bibinfo {author} {\bibfnamefont {D.}~\bibnamefont
  {Budker}}\ and\ \bibinfo {author} {\bibfnamefont {M.}~\bibnamefont
  {Romalis}},\ }\href {http://dx.doi.org/10.1038/nphys566} {\bibfield
  {journal} {\bibinfo  {journal} {Nat Phys}\ }\textbf {\bibinfo {volume} {3}},\
  \bibinfo {pages} {227} (\bibinfo {year} {2007})}\BibitemShut {NoStop}%
\bibitem [{\citenamefont {Pustelny}\ \emph {et~al.}(2008)\citenamefont
  {Pustelny}, \citenamefont {Wojciechowski}, \citenamefont {Gring},
  \citenamefont {Kotyrba}, \citenamefont {Zachorowski},\ and\ \citenamefont
  {Gawlik}}]{PustelnyJAP2008}%
  \BibitemOpen
  \bibfield  {author} {\bibinfo {author} {\bibfnamefont {S.}~\bibnamefont
  {Pustelny}}, \bibinfo {author} {\bibfnamefont {A.}~\bibnamefont
  {Wojciechowski}}, \bibinfo {author} {\bibfnamefont {M.}~\bibnamefont
  {Gring}}, \bibinfo {author} {\bibfnamefont {M.}~\bibnamefont {Kotyrba}},
  \bibinfo {author} {\bibfnamefont {J.}~\bibnamefont {Zachorowski}}, \ and\
  \bibinfo {author} {\bibfnamefont {W.}~\bibnamefont {Gawlik}},\ }\href
  {http://scitation.aip.org/content/aip/journal/jap/103/6/10.1063/1.2844494}
  {\bibfield  {journal} {\bibinfo  {journal} {Journal of Applied Physics}\
  }\textbf {\bibinfo {volume} {103}},\ \bibinfo {eid} {063108} (\bibinfo {year}
  {2008})}\BibitemShut {NoStop}%
\bibitem [{\citenamefont {Horrom}\ \emph {et~al.}(2012)\citenamefont {Horrom},
  \citenamefont {Singh}, \citenamefont {Dowling},\ and\ \citenamefont
  {Mikhailov}}]{HorromPRA2012}%
  \BibitemOpen
  \bibfield  {author} {\bibinfo {author} {\bibfnamefont {T.}~\bibnamefont
  {Horrom}}, \bibinfo {author} {\bibfnamefont {R.}~\bibnamefont {Singh}},
  \bibinfo {author} {\bibfnamefont {J.~P.}\ \bibnamefont {Dowling}}, \ and\
  \bibinfo {author} {\bibfnamefont {E.~E.}\ \bibnamefont {Mikhailov}},\ }\href
  {\doibase 10.1103/PhysRevA.86.023803} {\bibfield  {journal} {\bibinfo
  {journal} {Phys. Rev. A}\ }\textbf {\bibinfo {volume} {86}},\ \bibinfo
  {pages} {023803} (\bibinfo {year} {2012})}\BibitemShut {NoStop}%
\bibitem [{\citenamefont {Kenney}\ and\ \citenamefont
  {Keeping}(1939)}]{KenneyBook1940}%
  \BibitemOpen
  \bibfield  {author} {\bibinfo {author} {\bibfnamefont {J.}~\bibnamefont
  {Kenney}}\ and\ \bibinfo {author} {\bibfnamefont {E.}~\bibnamefont
  {Keeping}},\ }\href {https://books.google.es/books?id=Jxx6nQEACAAJ} {\emph
  {\bibinfo {title} {Mathematics of statistics part II}}},\ \bibinfo {edition}
  {2nd}\ ed.\ (\bibinfo  {publisher} {Chapman and Hall, London},\ \bibinfo
  {year} {1939})\BibitemShut {NoStop}%
\end{thebibliography}%

\clearpage

\newcommand{\TITLE}{Sensitivity, quantum limits,  and quantum enhancement of noise spectroscopies}

\onecolumngrid
{\Large Supplementary Material  for \textit{\TITLE} by Lucivero \textit{et al.}  \\ ~ \\ }
%\end{widetext}
\twocolumngrid
\section{statistics of $S$. }

We first note that if $Y(t)$ is a stationary gaussian random process, so that $P(Y(t_1 +\tau) = y_1, Y(t_2 +\tau ) = y_2, \ldots , Y(t_n+\tau) = y_n)$ is normally distributed and independent of $\tau$, then its Fourier transform $\tilde{Y}(\nu_j) \equiv \sum_m \exp[i \nu_j t_m] Y(t_m)$ (for brevity we omit normalization) has the following properties.  Writing $\tilde{Y}(\nu_j) = x_j + i y_j = r_j \exp[i \theta_j]$ with real $x,y,r$ and $\theta \in [0,2 \pi)$, linearity of the sum implies that $x_j$ and $y_j$ are normally distributed. For $\nu_j \ne 0$, stationarity implies $\tilde{Y}(\nu_j) = \sum_m \exp[i \nu_j (t_m + \tau)] Y(t_m) = \exp[i \nu_j \tau]  \tilde{Y}(\nu_j)$, independently of the value of $\tau$. As a consequence, $\theta_j$ must be uniformly distributed on $[0,2 \pi)$. This phase invariance implies equal distributions for $x$ and $y$, and that $r_j^2/\sigma_x^2 = (x_j^2 + y_j^2)/\sigma_x^2$ is described by the chi-squared distribution $\chi^2_k$ with $k=2$, since $x_j/\sigma_x$ and  $y_j/\sigma_x$ are independent unit-variance normal random variables. By well-known properties of the $\chi^2$ distribution, $\var(r_j^2) = \langle r_j^2 \rangle^2$ and thus $\var[S(\nu_j)] = \langle S(\nu_j) \rangle^2$.

\section{Alternative derivation of Eq. (3)}

Here we give a proof of Eq. (3) using the error propagation formula, which may provide insights not evident in the more abstract treatment using Fisher information. As in the main text, the spectrum is described by a multivariate normal distribution with mean $\langle \bar{S}_i \rangle =  f(\nu_i, {\bf v}) \equiv f_i$, variance
%$\sigma_i^2 = \var( \bar{S}_i ) =  f_i/{\cal N}$
$\var( \bar{S}_i ) =  f_i/{\cal N}$
and cross-correlations $\cov(\bar{S}_i, \bar{S}_j) = 0$  for $i \ne j$.
Maximum likelihood estimation is performed by minimizing  $\chi^2 \equiv \sum_i (f_i-\bar{S}_i)^2/(\BinxAve f_i)$, i.e., finding $\hat{\bf v}$ such that
\begin{equation}
\left. \partial_j \chi^2(\mathbf{v})\right|_{\mathbf{v} = \hat{\mathbf{v}}}  =  {0}
\end{equation}
where $\partial_j$ indicates $\partial/\partial v_j$ and $v_j$ is a component of the parameter vector $\mathbf{v}$.
%The carat over $\hat{\mathbf{v}}$ indicates the estimator of $\mathbf{v}$.
%From here on, we suppress the $\mathbf{v}$ dependence in $f_i(\mathbf{v})$ and we use the Einstein summation convention.
Applying the derivative, the optimization condition can be written (we use the Einstein summation convention from here forward)
%\begin{equation}
%\label{eq:Fit}
%2 \left(0 + \frac{S_i}{f_i^2} - \frac{S_i^2}{f_i^3}
%\right) \partial_j f_i = 0
%\end{equation}
%\begin{equation}
%\label{eq:Fit}
% \left[
%\left(\frac{S_i}{f_i^2} - \frac{S_i^2}{f_i^3}
%\right) {\partial_j f_i}{} \right]_{\mathbf{v} = \hat{\mathbf{v}}} = 0
%\end{equation}
%taking the total derivative
%\begin{eqnarray}
%\label{eq:Fit}
%\left. \partial_j \chi^2(\mathbf{v})\right|_{\mathbf{v} = \hat{\mathbf{v}}}  & = &
%\left[
%\delta S_i \left( \frac{1}{f_i^2} - \frac{2 S_i}{f_i^3} \right) \partial_j f_i
%\right. \nonumber \\ & & + \left.
%  \delta v_k  \left( \frac{-2 S_i}{f_i^3} - \frac{6 S_i^2}{f_i^4} \right) (\partial_k f_i) \partial_j f_i
%  \right. \nonumber \\ & & + \left.
%   \delta v_k \left(\frac{S_i}{f_i^2} - \frac{S_i^2}{f_i^3}
%\right) \partial_k \partial_j f_i
%\right]_{\mathbf{v} = \hat{\mathbf{v}}} = 0
%\end{eqnarray}
\begin{equation}
\label{eq:Fit}
\left.\frac{(f_i-\bar{S}_i)\partial_jf_i}{f_i^2}\right|_{\mathbf{v} = \hat{\mathbf{v}}} = \left.\frac{(f_i-\bar{S}_i)^2 \partial_jf_i}{f_i^3}\right|_{\mathbf{v} = \hat{\mathbf{v}}}.
\end{equation}
We can apply the variational principle to understand how a small error in $\bar{S}_i$ i.e. $\bar{S}_i=f_i+\delta \bar{S}_i$ produces a corresponding error in the estimator $\hat{v}_k={v}_k+\delta v_k$.
%By assuming that the fit function correctly describes the mean spectrum, i.e. $f-\Sbar = 0$, to
The r.h.s. of  Eq. (\ref{eq:Fit})  is of order $(\delta \bar{S}_i)^2$ and thus vanishingly small compared to the l.h.s. in the asymptotic limit. The l.h.s. similarly vanishes, maintaining the optimum, provided that
\begin{equation}
\label{eq:varprinciple}
\frac{(\partial_kf_i)(\partial_j f_i)}{f_i^2}\delta\hat{v}_k=\frac{\partial_jf_i}{f_i^2}\delta \bar{S}_i
\end{equation}
%which is a general result for a noise spectrum $S(\nu)$ that is fitted with the right model $f_i=\Sbar_i$ and that satisfies Eq. (1) of the main paper.
At this point it is convenient to introduce the matrices
\begin{equation}
L_{ij} \equiv \frac{\partial_j f_i}{f_i}
\end{equation}
and
\begin{equation}
M_{jk} \equiv  L_{ji} L_{ki}  = \frac{(\partial_kf_i)(\partial_j f_i)}{f_i^2}
\end{equation}
We note the relation to the Fisher information of Eq. (3): ${\cal I} = ({\cal N} + 2) M$.
% In order to obtain an analytical expression for the covariance matrix $\Gamma$ of the estimator $\hat{\mathbf{v}}$, we define a non-square matrix $L$:
%\begin{equation}
%L_{ij} \equiv \frac{\partial_j f_i}{f_i}.
%\end{equation}
%(no summation) and, by defining the matrix $M \equiv L L^T$, i.e. $M_{jk} \equiv  L_{ji} L_{ki}$, we can rewrite
Equation (\ref{eq:varprinciple}) then becomes
\begin{equation}
M_{jk}  \delta \hat{v}_k = L_{ji} \frac{\delta \bar{S}_i}{f_i}
\end{equation}
with solution
\begin{equation}
\delta \hat{v}_k =  M_{kj}^{-1} L_{ji} \frac{\delta \bar{S}_i}{f_i}.
\label{eq:deltav}
\end{equation}
%, which has been directly reported in the main text.
%Eq. (\ref{eq:deltav}) relates the uncertainty of the spectrum, i.e. the noise on the noise power, with the uncertainty on the fit parameters.
The elements of the covariance matrix of $\hat{\mathbf{v}}$ are given by:
\begin{eqnarray}
\Gamma_{jk} &\equiv& \langle\hat{v_j}\hat{v}_k\rangle - \langle\hat{v_j}\rangle\langle\hat{v}_k\rangle = \langle\delta\hat{v_j}\delta\hat{v}_k\rangle \nonumber \\
&=& \Big\langle\frac{M_{jl}^{-1} L_{lm}\delta \bar{S}_m(M_{kn}^{-1} L_{np}\delta \bar{S}_p)}{f_mf_p}\Big\rangle \nonumber \\
&=& \Big\langle M_{jl}^{-1} L_{lm}L_{pn}M_{nk}^{-1}\frac{\delta \bar{S}_m\delta \bar{S}_p}{f_mf_p}\Big\rangle
\label{eq:CovMat}
\end{eqnarray}
Using
\begin{eqnarray}
{\langle \delta \bar{S}_m \delta \bar{S}_p \rangle} & = & {f_m f_p} \delta_{mp} / {\cal N}
\end{eqnarray}
we obtain:
\begin{eqnarray}
\Gamma_{jk} &=& M_{jl}^{-1} L_{lm}L_{pn}M_{nk}^{-1}\delta_{mp} {\cal N}^{-1} \nonumber \\
%&=& M_{jl}^{-1} L_{lm}L_{mn}M_{nk}^{-1}  {\cal N}^{-1}\nonumber \\
%&=& M_{jl}^{-1}M_{ln}^{}M_{nk}^{-1} {\cal N}^{-1} \nonumber \\
&=& M_{jk}^{-1} {\cal N}^{-1}
\label{eq:CovMat2}
\end{eqnarray}
which approaches Eq. (3) of the main text for large ${\cal N}$. %r $({\cal N} + 2)/{\cal N} \rightarrow 1$ for large ${\cal N}$.

\section{Spectral parameters versus optical power and atomic density}
\label{sec:expr}
As described in detail in \cite{Lucivero2016}, the parameters $(S_{\rm{ph}},\nu_L,S_{\rm{at}},\Delta \nu)$ can be computed from first principles.  The contribution from photon shot-noise to the spin noise spectrum $S(\nu)$ is given by:
\begin{equation}
   S_{\rm ph}  = 2G^2 q (\Re P) \xi^{2},
    \label{Eq:Sph}
\end{equation}
\noindent where $\xi^2$ is the squeezing factor, $G =\unit{10^{6}}{ \volt \per {\ampere}}$ is the transimpedance gain, $P$ is the total optical power reaching the detector, $\Re =  \eta q/\Eph$ is the detector responsivity, $\eta$ denotes the quantum-efficiency of the detector,  $\Eph= \hbar \omegalight = \unit{2.49\times 10^{-19}}{\joule}$ is the photon energy at \unit{795}{nm}, and $q = \unit{1.6\times 10^{-19}}{C}$ is the electron charge. The on-resonance contribution from the atomic $^{85}$Rb spin noise is given by:
\begin{equation}
    S_{\rm at}  = \frac{8G^2(\Re P)^2\kappa^2 \Aeff \Lcell  n^{(87)}}{\pi \Delta \nu}
    \label{Eq:Sat}
\end{equation}
where $\Aeff=0.054$ cm$^2$ is the effective area, $\Lcell=3$ cm is the vapor cell length and $\kappa^2$ is a parameter including the spectral factor, defined in Eq. (13) of \cite{Lucivero2016}. $n^{(87)} = 0.72 n$ is the number density of $^{85}$Rb at natural abundance.  The FHWM linewidth is given by
\begin{equation}
   \Delta \nu  = \frac{1}{T_2 \pi} = \frac{1}{\pi}( \Gamma_0 +\alpha n +\beta P)
    \label{Eq:FWHM}
\end{equation}
where $\Gamma_0$ is the unperturbed relaxation rate, $\alpha$ and $\beta$ are collisional and power broadening factors, respectively. Finally, the Larmor frequency $\nu_L = \gamma B$, where $\gamma$ is the gyromagnetic ratio and $B$ is the magnetic field strength.  In our range of parameters, $\nu_L$ is not significantly affected by light shifts or collisional shifts and is taken as constant. Numerical values of the defined variables, for our experimental conditions, could be found in Table I in \cite{Lucivero2016}.

\section{Sensitivity to shot noise level $\Gamma^{\rm th}_{11}(n,P)$ and  peak atomic noise $\Gamma^{\rm th}_{33}(n,P)$}

In Fig. (\ref{Fig:ContourPlotsg11g33}) we show $2$D contour-plots of the variances of the estimated atomic $S_{\rm{at}}$ and shot noise $S_{\rm{ph}}$ contributions to the power spectrum, given by the covariance matrix diagonal terms $\Gamma^{\rm th}_{33}(n,P)$ and $\Gamma^{\rm th}_{11}(n,P)$, respectively.

\begin{figure}[h!]
\centering
\includegraphics[width=0.49 \columnwidth]{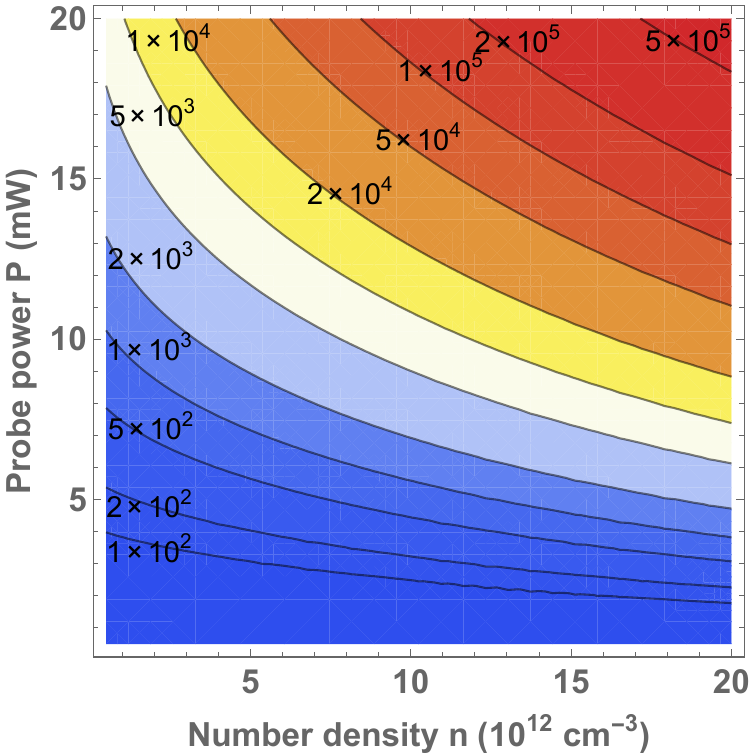}
\includegraphics[width=0.49 \columnwidth]{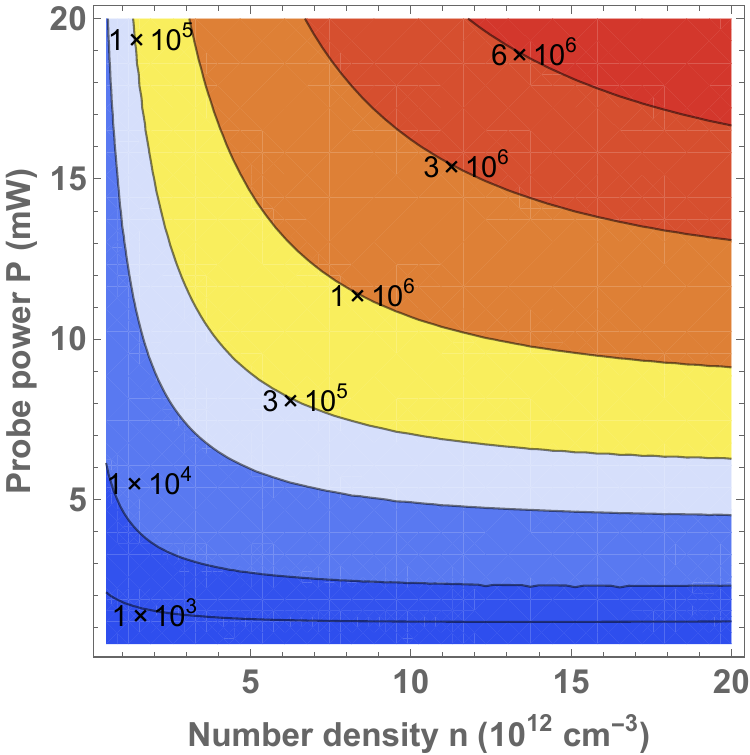}
  \caption{ $\GamTh_{11}$ (left), variance in shot noise power spectral density and $\GamTh_{33}$ (right), variance in peak spin noise power spectral density,  as predicted by Eq.~(3) as a function of number density and probe power.  Both in units of \unit{}{\micro\volt^{4}\per\hertz\squared}.
}
%
%  \caption[2D Contour plots of covariance matrix terms $\Gamma^{\rm th}_{33}(n,P)$ and $\Gamma^{\rm th}_{11}(n,P)$]{\textbf{Analytical results.} \textbf{(Left)} Covariance matrix term $\Gamma^{\rm th}_{33}(n,P)$ versus atomic density and optical probe power. \textbf{(Right)} Covariance matrix term $\Gamma^{\rm th}_{11}(n,P)$ in the same parameters space. In the paper sample data are shown for $P=2$ mW (continuous horizontal red line) and $n=7.65*10^{12}$ cm$^{-3}$ (continuous vertical black line) (see text) \mtext{fix caption}}
  \label{Fig:ContourPlotsg11g33}
\end{figure}

For these fit parameters the theoretical variance increases monotonically with both power and density, without showing an optimal region or an inversion trend within the investigated parameter range, differently from the variance of the estimated Larmor frequency $\nu_L$ and resonance linewidth $\Delta\nu$, as shown in the main paper.

%\section{Standard error on the sample covariance matrix}
%The standard deviation $\sigma^{\rm th}$ is calculated by knowing that the sample covariance matrix $\Gamma$ of a sample from a multivariate normal distribution follows the Wishart probability density function \cite{Wishart1928}:
%\begin{equation}
%P(\Gamma^{\rm th})=\det\Gamma^{(m-p-1)/2}e^{-\frac{{\rm Tr}[\Gamma^{\rm th}\Gamma]}{2}}
%\label{eq:Wishart}
%\end{equation}
%with sample size $m=100$, number of variables $p=4$ for the our sample data analysis, mean value $\Gamma^{\rm th}$ and variance ${\rm var}(\Gamma^{\rm th})=(\sigma^{\rm th})^2$, where for ${\rm var}(M)$ of a generic matrix $M$ we mean the matrix with elements $[{\rm var}(M)]_{ij}={\rm var}(M_{ij})$ and for $M^2$ we mean the matrix with elements $[(M)^2]_{ij}=M_{ij}^2$.

~

\section{Calculation of experimental variances and their uncertainties}

The variances from experimental observations are computed as $\Gamma_{ii}^{\rm exp} = k_2$, the second ``k-statistic,'' i.e., Fisher's unbiased estimator for the second cumulant.  $k_2$ and  $k_4$, fourth k-statistic, are computed from $m$ observations $v^{(1)}, \ldots , v^{(m)} $ as
\begin{equation}
k_2=\frac{m(S_2-S_1^2)}{m(m-1)}
\label{eq:k2}
\end{equation}
and
\begin{eqnarray}
k_4&=&\frac{-6S_1^4 + 12mS_1^2S_2 -3m(m-1)S_2^2}{m(m-1)(m-2)(m-3)}\nonumber\\
   &+&\frac{-4m(m+1)S_1S_3+m^2(m+1)S_4}{m(m-1)(m-2)(m-3)}
\label{eq:k4}
\end{eqnarray}
where
\begin{equation}
S_r\equiv\sum_{i=1}^{m}(v^{(i)})^r.
\label{eq:kpowers}
\end{equation}
The estimated variance in $k_2$ is then calculated using the unbiased estimator for $\var(k_2)$  \cite[pp. 189-190]{KenneyBook1940} :
\begin{equation}
{\rm var}(\Gamma_{ii}^{\rm exp})= {\rm var}(k_2)=\frac{2nk_2^2+(n-1)k_4}{n(n+1)}.
\label{eq:vark2}
\end{equation}
%Then, an unbiased estimator for ${\rm var}(k_2)$, which is the variance of the estimator of the variance of the distribution, is given by (\cite{KenneyKeeping} pp. 189-190):
%\begin{equation}
%{\rm var}(\Gamma_{ii}^{\rm exp})=
%%{\rm var}(k_2)=
%\frac{2nk_2^2+(n-1)k_4}{n(n+1)}
%\label{eq:vark2}
%\end{equation}
From Eq. (\ref{eq:vark2}) we obtain the standard error of the diagonal elements of the sample covariance matrix $\Gamma_{ii}^{\rm exp}$.
%By performing this calculation with $i=1$ to $4$ we get the error bars on data that are shown in the main paper.

\end{document}